\documentclass[journal,twoside]{Definitions/IEEEtran}
\usepackage{cite}

\ifCLASSINFOpdf
   \usepackage[pdftex]{graphicx}
  \graphicspath{{../Images/}{../jpeg/}}
  \DeclareGraphicsExtensions{.pdf,.jpeg,.png}
\else
\fi
\usepackage{amsmath}
\usepackage{algorithm}
\usepackage[noend]{algpseudocode}
\usepackage{multicol}

\usepackage{array}

\usepackage{mdwmath}
\usepackage{mdwtab}

\ifCLASSOPTIONcompsoc
  \usepackage[caption=false,font=normalsize,labelfont=sf,textfont=sf]{subfig}
\else
  \usepackage[caption=false,font=footnotesize]{subfig}
\fi

\usepackage{booktabs}
\usepackage[normalem]{ulem}
\useunder{\uline}{\ul}{}
\usepackage{diagbox}
\usepackage{tabularx}
\usepackage{rotating}
\usepackage{enumerate}
\usepackage{dirtytalk}

\usepackage{scalerel}
\usepackage{tikz}
\usetikzlibrary{svg.path}

\definecolor{orcidlogocol}{HTML}{A6CE39}
\tikzset{
  orcidlogo/.pic={
    \fill[orcidlogocol] svg{M256,128c0,70.7-57.3,128-128,128C57.3,256,0,198.7,0,128C0,57.3,57.3,0,128,0C198.7,0,256,57.3,256,128z};
    \fill[white] svg{M86.3,186.2H70.9V79.1h15.4v48.4V186.2z}
                 svg{M108.9,79.1h41.6c39.6,0,57,28.3,57,53.6c0,27.5-21.5,53.6-56.8,53.6h-41.8V79.1z M124.3,172.4h24.5c34.9,0,42.9-26.5,42.9-39.7c0-21.5-13.7-39.7-43.7-39.7h-23.7V172.4z}
                 svg{M88.7,56.8c0,5.5-4.5,10.1-10.1,10.1c-5.6,0-10.1-4.6-10.1-10.1c0-5.6,4.5-10.1,10.1-10.1C84.2,46.7,88.7,51.3,88.7,56.8z};
  }
}

\newcommand\orcidicon[1]{\href{https://orcid.org/#1}{\mbox{\scalerel*{
\begin{tikzpicture}[yscale=-1,transform shape]
\pic{orcidlogo};
\end{tikzpicture}
}{|}}}}

\usepackage{hyperref} 

\begin{document}
%
\title{A General Purpose Data and Query Privacy Preserving Protocol for Wireless Sensor Networks}
%
%
%

\author{Niki Hrovatin \orcidicon{0000-0003-1082-0697},
        Aleksandar Tošić \orcidicon{0000-0001-5627-4420},
        Michael Mrissa \orcidicon{0000-0002-2330-1004},
        and~Jernej Vičič \orcidicon{0000-0002-7876-5009}
\thanks{Manuscript received October 15,2021. The authors gratefully acknowledge the European Commission for funding the InnoRenew project (Grant Agreement \#739574) under the Horizon2020 Widespread-Teaming program and the Republic of Slovenia (Investment funding of the Republic of Slovenia and the European Regional Development Fund). They also acknowledge the Slovenian Research Agency ARRS for funding the project J2-2504. (Corresponding author: Niki Hrovatin)}
\thanks{Niki Hrovatin, Aleksandar Tošić and Michael Mrissa are with InnoRenew CoE, Livade 6, 6310 Izola, Slovenia, and also with University of Primorska, Faculty of Mathematics, Natural Sciences and Information Technologies, Glagoljaška 8, SI-6000 Koper, Slovenia (e-mail: niki.hrovatin@famnit.upr.si).}
\thanks{Jernej Vičič is with the Research Centre of the Slovenian Academy of Sciences and Arts, The Fran Ramovš Institute, Novi trg 2, 1000 Ljubljana, Slovenia and also with University of Primorska, Faculty of Mathematics, Natural Sciences and Information Technologies, Glagoljaška 8, SI-6000 Koper, Slovenia.}
}

\maketitle

\begin{abstract}

Wireless Sensor Networks (WSNs) are composed of a large number of spatially distributed devices equipped with sensing technology and interlinked via radio signaling. A WSN deployed for monitoring purposes can provide a ubiquitous view over the monitored environment. However, the management of collected data is very resource-consuming and raises security and privacy issues.
In this paper, we propose a privacy preserving protocol for collecting aggregated data from WSNs. The protocol relies on the Onion Routing technique to provide uniformly distributed network traffic and confine the knowledge a foreign actor can gain from monitoring messages traveling the network.
Our solution employs the computing power of nodes in the network by conveying them general-purpose computer code for in-situ processing and aggregation of data sourcing from multiple sensor nodes. 
We complement our work with a simulation of the proposed solution using the network simulator ns-3. Results of the simulation give an overview of the scalability of the solution and highlight potential constraints.


\end{abstract}

\begin{IEEEkeywords}
Data Aggregation, Edge Computing, ns-3, Onion Routing, Privacy, Wireless Sensor Networks.
\end{IEEEkeywords}

%
\IEEEpeerreviewmaketitle

\section{Introduction}
\label{sec:introduction}
%
%
%
%
\IEEEPARstart{O}{ver the last} few years, the cost reduction of sensor production and wireless technologies have contributed to the development of large scale Wireless Sensor Networks (WSN).
Nowadays, WSNs are composed of dozens or hundreds of sensing nodes interlinked via radio signaling and meant to be easily deployed, self-configurable and low cost. Those nodes are sensing and reporting environmental data to other nodes dedicated to data collection, called sink nodes.
In a typical WSN data is moving through a wireless multi-hop network without infrastructure. 
Even though wireless communication is typically secured by encryption, low-cost devices, multi-hop routing, and the lack of infrastructure make WSNs subject to various attacks. 
The research~\cite{chan2003security} states that the WSN technology will benefit our daily lives in important ways and that: \say{We cannot deploy such a critical technology, however, without first addressing the security and privacy research challenges to ensure that it does not turn against those whom it is meant to benefit.}

The large number of sensing nodes composing a WSN allows very granular monitoring of the environment; however, the large amount of data produced can overload the resource-constrained technology WSNs consist of. Particularly, sending unprocessed data causes unnecessary communication overhead and high computation load over the sink nodes. Moreover, the raw data collected is often redundant since the sensing range of neighboring nodes is frequently overlapping. 

Therefore, the recently emerging edge computing paradigm raises the notion of moving computations as close as possible to data sources (i.e., sensor nodes) in order to alleviate the network and to take advantage of the increased computing power of sensor nodes.
Only significant, already processed data is sent over the network.


While encryption methods can adequately protect the data itself, a wireless network populated by exclusively essential traffic could still reveal many details to an actor eavesdropping the wireless communication. The specific nature of traffic could reveal information about what kind of activity is taking place in the monitored environment; moreover, the path that data traverses from the data source node towards the sink node can reveal the exact location of the monitored activity. 
Recent studies show that with traffic analysis it is possible to extract features like message size, frequency, processing time, and associating these features with facts or secrets makes machine learning techniques able to infer over current activities in the monitored region~\cite{gu2020iotspy,zhang2011defending,saltaformaggio2016eavesdropping}.

Drawing upon the discussed privacy needs, we aim at designing a solution for collecting data from a WSN that meets the following requirements:
\begin{enumerate}
    \item \textit{Privacy preservation against traffic analysis:} The communication traffic in the WSN must not leak information of the monitored environment to external actors eavesdropping on the wireless communications.
    \item \textit{Privacy preservation against internal threats:} 
    The solution must conceal the relation between a sensor node in the network and a data collection operation to avoid disclosing information on the data collection. When collecting data, the private sensed values of a sensor node must not be disclosed to any other sensor node in the network. Moreover, disclosing these details must be hard even if multiple sensor nodes are maliciously collaborating. 
\end{enumerate}

This paper addresses the defined requirements with a protocol designed to query a subset of sensor nodes in a WSN while preserving data and query privacy. 
The need for such a solution is motivated by WSNs for building monitoring~\cite{clements2011sustainable,ghayvat2015enhancement,airQ}. 
In particular, we are looking into the specific case of indoor air quality monitoring since it receives an increasing interest as it contributes to reducing the environmental impact of buildings, to improving building occupants' well-being, and to enhancing future building design~\cite{airQ}. Although indoor air quality monitoring systems based on WSNs provide many benefits due to their granular monitoring capabilities, the broadcast nature of wireless communication makes them attractive for attackers with malicious intentions, such as compromising building security.

Our contribution is structured as follows:

\begin{enumerate}
    \item We propose a privacy preserving querying protocol for collecting aggregated data from a WSN. Our protocol harnesses the computing power of sensors nodes on the edge of the network to realize the execution of arbitrary general-purpose computer code dedicated to data collection. Indeed, data aggregation takes place on sensor nodes while computation and intermediate aggregation results move across the network.
    \item We support our work with analyses meant to identify vulnerabilities concerning privacy preservation. First, we analyze how uniform message size, encryption, query forwarding time decoupled from query execution, randomized forwarding paths, and the inclusion of decoy nodes imitating query execution make the protocol secure against actors listening to wireless communications. Second, the analysis delves into privacy concerns over trusted nodes that passively collaborate in disclosing sensed values of other nodes. Precisely the protocol takes advantage of the onion routing technique to route a message through a strictly defined path and to secretly deliver encryption keys only to specific nodes in the message path. Encryption keys are essential to access the message body containing the computer code and the aggregated data. 
    \item We developed a simulation of the proposed protocol using the network simulator ns-3~\cite{henderson2008network,ns3}. The simulation was used to run two experiments:
    \textit{a)} examine how the solution scales in large WSNs; \textit{b)} determine if the arrangement of nodes in the environment affects the functioning of the protocol.  
\end{enumerate}

The rest of this paper is organized as follows:
Section~\ref{sec:scenario} characterizes the motivating scenario for our work.
Section~\ref{sec:solution_overview} gives a brief overview of the privacy preserving protocol.
Section~\ref{sec:related} reviews related work and highlights the originality of our solution.
Section~\ref{sec:research_contribution} details the general-purpose data and query privacy preserving protocol.
Section~\ref{sec:analyses} gives analyses concerning privacy preservation.
Section~\ref{sec:results} presents results of the privacy preserving communication protocol simulated using the network simulator ns-3.
Section~\ref{sec:discussion} discusses the findings and results of Section~\ref{sec:analyses} and Section~\ref{sec:results}.
Section~\ref{sec:conclusion} concludes the manuscript and gives guidelines for future work.

\section{Motivating Scenario}
\label{sec:scenario}

In this paper, we motivate our work with a building monitoring scenario that relies on a WSN equipped with
sensors collecting data about temperature, $CO_2$, $VOC$, $PM$, relative humidity, etc.
However, a large WSN relying on high quality sensors for air quality monitoring can reach prohibitive costs.
A common practice to reduce the overall WSN cost is the dense deployment of nodes equipped with low-cost sensor technology. 
Since neighboring nodes of a densely deployed WSN often share the same monitoring area, it is possible to improve reading quality by aggregating data from multiple low-cost sensors.
Indeed, a room equipped with multiple sensors will provide a better view of the monitored environment than a single sensor, as, for example, particulate matter sensors are affected by activities happening in their proximity, like cooking, cleaning, moving objects, etc.
Hence, aggregating data from multiple sensors can reduce the number of triggered false alarms.
However, in typical WSN implementations, the data is aggregated only at network endpoints resulting in high communication overhead since each node needs to report its sensed value and missing the opportunity to take advantage of the nodes' computing capabilities when conveying the whole processing load on the end system.
Additionally, although indoor air quality monitoring is typically implemented with WSNs protected with Secure Socket Layer (SSL) to encrypt data, such design makes network endpoints vulnerable to attacks such as traffic analysis.
Indeed, external actors can observe communication patterns and node activities to gain knowledge over the network topology, and unusual changes in traffic patterns could be used to infer about the state of an area where the WSN is located. 
Moreover, the network traffic could reveal even more information if sensor nodes are not all sensing the same environmental features.

The described scenario motivates the need to combine privacy preservation and distributed data computing on WSN nodes to employ node's computing capabilities without revealing sensor readings or sensitive contextual information that can disclose the sensors equipped onto nodes or the need for a specific data request.

\begin{figure*}[th]
\centering
\includegraphics[scale=0.50]{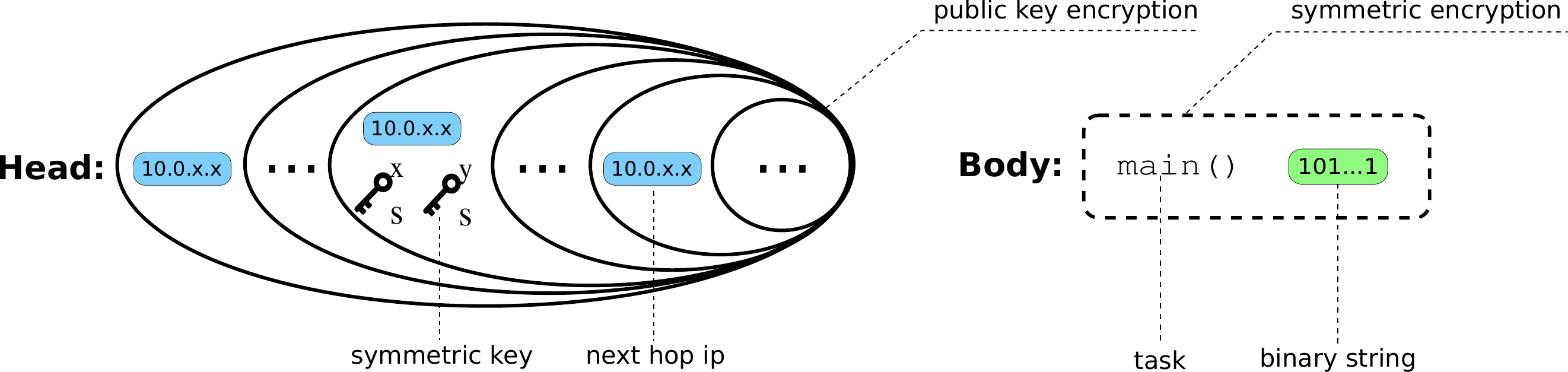}
 \caption{Representation of the query, made of the head (left) and the body (right).}
 \label{fig:querEx}
\end{figure*}

\begin{figure*}[h]
\centering
\includegraphics[scale=0.55]{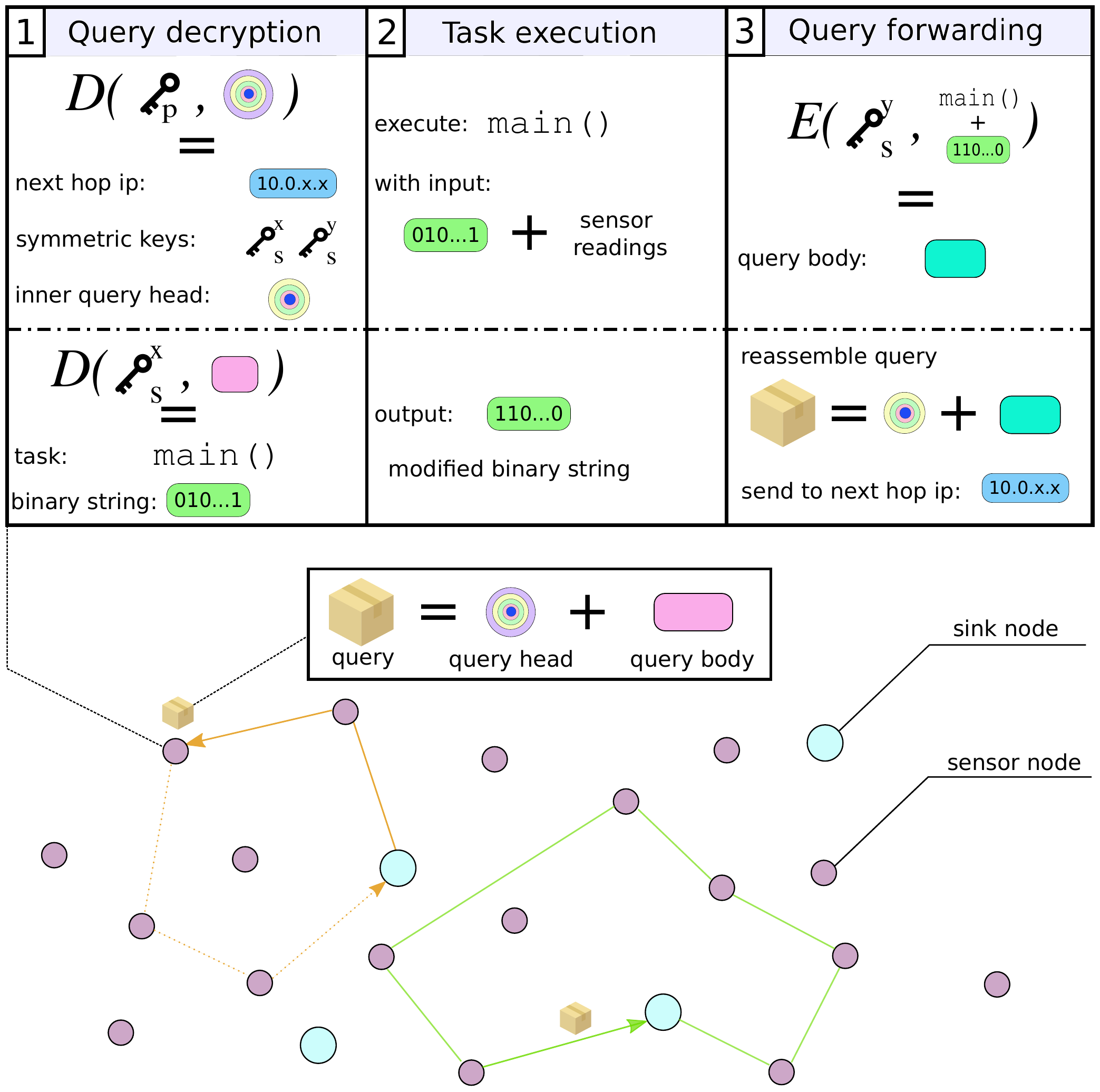}
 \caption{Overview of query processing at a target sensor node in three phases (\textit{query decryption, task execution, query forwarding}). Moreover, the figure displays the query's path forming a circuit.}
 \label{fig:querying}
\end{figure*}

\section{Solution overview}
\label{sec:solution_overview}

To address the previously described scenario, we propose a communication protocol based on the Onion Routing~\cite{syverson1997anonymous} technique for anonymous communication over a computer network.
We similarly employ messages structured into encryption layers, such that a layer can be decrypted only by the targeted node revealing an inner encryption layer addressed to another node in the network.
Therefore, message decryption is carried out gradually by leading the message across WSN nodes following the precise order given at message construction.
Encryption layers are not enclosing only the inner layer, but also additional secret information revealed only to the node decrypting that layer.
The proposed solution takes advantage of this peculiarity to convey path details and symmetric encryption keys to WSN nodes. 
We deliver path details in encryption layers to prevent the disclosure of the whole message path, such that a node receiving the message can identify only the sender and the next receiver of the message. Encryption keys, however, are delivered only to a subset of nodes in the message path. 
Moreover, and differently from onion routing, the layered object is accompanied with information related to edge data processing revealed only to those nodes receiving symmetric encryption keys.
Therefore, nodes in the message path serve as an anonymity set\footnote{Based on the definition given by Pfitzmann and K\"{o}hntopp\cite{pfitzmann2010terminology}, the anonymity set is the set of subjects that might cause an action. In our solution, the anonymity set is the set of nodes deciphering a layer of a layered object. If layer decryption reveals symmetric encryption keys to a node, then the node executes edge data processing; the action.} for nodes executing the edge data processing since each node in the message path could potentially receive symmetric encryption keys from the decryption of the layered object.

This paper describes a technique to retrieve aggregated data from a WSN that acts as a service. Indeed, external actors pose requests to sink nodes serving as the gateway to access the WSN. A sink node in charge of a request will construct a query. We refer to a query as the message composed of a head consisting of the previously described message of encryption layers and a body consisting of the pair \emph{$<$task, binary string$>$} as shown if Fig.~\ref{fig:querEx}. We refer to a task as the computer code sent in execution to sensor nodes. The binary string is of fixed size and carries task execution results back to the sink node that issued the query since each query head is constructed to lead the query over a path forming a circuit.
We refer to the path of a query as the nodes traversed by the query that decrypt one of the query head layers.
The query body is encrypted using the symmetric encryption key present in the query head precisely in the outermost encryption layer holding encryption keys. 
Query processing is depicted in Fig.~\ref{fig:querying} and explained in the next paragraph.

When a node in the query path receives the query, it decrypts the outermost encryption layer of the query head, revealing the next-hop address (IP) or the next-hop address and symmetric keys. A node receiving the query will behave differently based on the content revealed from the query head decryption.
\begin{enumerate}
    \item Nodes that obtain only the next-hop address are decoy nodes that simulate query processing by holding the query for a randomly chosen amount of time. The query head is padded to its original size, and the query is forwarded to the node at the next-hop address.
    \item Nodes that obtain both next-hop address and symmetric keys from query head decryption are nodes target of the query. A target node decrypts the query body using the first symmetric key obtained from query head decryption. Thus the query body reveals the pair \emph{$<$task, binary string$>$}. The task is executed in an execution environment that provides restricted access to the underlying sensor node system; specifically, the task must have access to sensor readings and the binary string. 
    Task execution results are embodied in the binary string. The query body is composed of the task and the modified binary string gets encrypted using the second symmetric encryption key obtained from query head decryption. The query head is padded to its original size, and after a randomly chosen amount of time, the query is forwarded to the node at the next-hop address.  
\end{enumerate}

The query's path ends at the sink node that issued the query. Therefore the sink node obtains the aggregated result from the binary string carried in the query body.

\section{Related Work}
\label{sec:related}

Many systems were developed to preserve users' privacy while communicating on large public networks like the Internet~\cite{ren2010survey}. Several solutions~\cite{gulcu1996mixing,sako1995receipt,berthold2001web} originate from the work of \textit{Chaum}~\cite{chaum1981untraceable} on mixnet. Mixnet-based schemes rely on a set of mix servers that receive encrypted messages, and after a sufficiently large amount of time, messages are re-ordered and released in batches to hide the correspondence between sender and receiver. Onion Routing~\cite{syverson1997anonymous} is a solution to preserve users' anonymity while avoiding latency introduced by mix servers. 
The solution relies on source routing using multiple encryption layers to route a message through a set of at least three routing servers (onion routers) 
to create an anonymous connection. 
Source routing~\cite{sunshine1977source} is a technique to route a message through a set of nodes by encoding path information in the message. In onion routing, path information is encrypted in each encryption layer of the message, and when the message is routed through onion routers, at each hop, a layer of encryption is removed from the message revealing the next hop. Therefore, no one of the actors involved in the communication will know the whole message path apart from the source of the message. In TOR~\cite{dingledine2004tor} the second-generation onion routing, the message path is similarly protected by encryption layers, but the anonymous connection is established incrementally using key exchange schemes. Although the mentioned solutions effectively provide anonymous communication between two parties, they rely on the background traffic of large networks. Furthermore, the mentioned solutions protect the logical location of communicating parties (IP-address), while in WSNs, the privacy of communicating nodes can also be disclosed by observing the physical wireless communication.

The literature review conducted by \textit{Li et al.} \cite{li2009privacy} divides the privacy problem in WSNs into data privacy and context privacy. 
Data privacy is achieved if a communication protocol does not leak insights about data to external and internal adversaries.
External adversaries can perform traffic analysis by eavesdropping on wireless communications.
In contrast, internal adversaries 
consist of nodes captured and controlled by malicious entities and may have knowledge of encryption keys used in the sensor network.
On the other hand, context privacy addresses concerns related to communication traffic characteristics, as this can reveal insights over monitored activities. 
Location privacy was extensively studied in event-driven WSNs using various routing strategies~\cite{jiang2019survey} aiming to conceal the location of data source nodes and sink nodes. The location of data source nodes can reveal insights over events detected by the WSN. On the other hand, keeping secret the location of sink nodes precludes the attacker from physically destroying sink nodes, which are of central importance for the correct functioning of the WSN. However, the application of similar techniques in our scenario is not adequate since we aim to aggregate data from multiple nodes that may be closely located, and routing strategies designed to anonymize source and destination might disclose the region of interest. 

The straightforward privacy-preserving approach to retrieve data from a region of interest in a WSN was highlighted by \textit{Carbunar et al.} \cite{carbunar2010query}, and consists in gathering data from all sensor nodes in the network and then keep readings only from the sensor nodes of interest. 
\textit{Xie et al.} \cite{xie2017efficient} proposed an efficient privacy-preserving compressive data gathering scheme. Compressive data gathering~\cite{xiang2012compressed} is based on Compressive Sensing~\cite{donoho2006compressed} a signal sampling technique that allows to reliably reconstruct the source signal from fewer samples than those required by the Nyquist theorem \cite{li2012compressed,wu2015efficient,middya2017compressive}. 
In compressive data gathering, each node participating in the data gathering will multiply its sensor reading with a vector of coefficients. The resulting vector is aggregated with other node vectors along the routing path to the sink node. The sink node can then recover raw sensory data from the aggregate. To introduce privacy preservation in the compressive data gathering scheme \textit{Xie et al.} proposed to employ public-key Homomorphic Encryption \cite{paillier1999public} to aggregate data along routing paths without decrypting intermediate aggregated results. The solution is also resilient to traffic analysis since messages containing aggregated values are changed by encryption after getting through nodes of the aggregation path. 

The solution efficiently gathers data from multiple sensor nodes. However, in WSNs, neighboring nodes often share the same monitored area, and data sensed from neighbor nodes is often correlated. Therefore, data gathering schemes designed to collect raw data are gathering many duplicated data, do not utilize sensor node computing capabilities to process data, and pose a substantial communication load on the network even if the data need involves a small subset of nodes of the network. 

In-network data aggregation \cite{intanagonwiwat2000directed,intanagonwiwat2002impact,tang2006extending} could effectively reduce the network's communication overhead and employ sensor node processing capabilities via the aggregation of multiple sensor node readings along routing paths toward the sink node. Moreover, the whole network works as a distributed processing mechanism delivering the final aggregated value to the sink node.

Privacy preserving data aggregation solutions ensure data privacy against external and internal attackers~\cite{conti2009privacy,zhang2008gp,zhang2013preserving,bista2010privacy}. However, data aggregation solutions do not address the problem of triggering a data aggregation process that affects only a subset of sensor nodes in the network without disclosing nodes participating in the aggregation or details about the aggregate to be computed.

In building air quality monitoring, the retrieval of aggregated data from the whole network of sensors can approximate the building air quality. However, obtaining data from individual locations of the building is imperative to give a granular assessment of the air quality. Therefore triggering a data aggregation process that affects a subset of nodes in the network is crucial in building monitoring scenarios. Furthermore, we aim to provide a general-purpose solution able to aggregate data from multiple sensor nodes satisfying specific conditions like exceeding a threshold value, fit a statistical measure, past sensed events, etc. If disclosed, such queries can reveal not only sensed data but also the region of interest, patterns of monitoring operations, the purpose and need for a specific query.

According to \textit{Carbunar et al.}~\cite{carbunar2010query} a privacy-preserving query mechanism in WSNs must hide from attackers the location and identity of queried sensor nodes but also the relationship between individual queries while maintaining an adequate trade-off between privacy and efficiency. \textit{Carbunar et al.} addressed query privacy needs with a WSN that acts as a service accessible through dedicated servers. The proposed solution hides query details from servers that provide access to the WSN. 
The WSN is mapped into regions, and queries are targeting individual regions. Query privacy is assured using source routing and by hiding a query constructed by the client with a number of additional bogus queries targeting different regions of the WSN.
However, the proposed solution does not address a privacy preserving query execution in the target region. 

\textit{De Cristofaro et al.} \cite{de2009privacy} proposed a privacy-preserving solution to retrieve individual sensor node readings without disclosing the identity of queried sensor nodes or data to the network owner or attackers. The solution relies on source routing \cite{sunshine1977source} using the onion routing technique to hide the query path and symmetric encryption to provide data privacy and data integrity. However, the proposed solution requires that the client issuing the query to the WSN knows the exact network topology to construct the source routing path.

The solution we are presenting diverges from the previously described approaches since we aim at retrieving aggregated data from the WSN by conveying arbitrary computer code through a sequence of sensor nodes. 
The onion routing is used to route the message through nodes and to permit only specific nodes to execute the computer code and thus contribute at the aggregate.
Furthermore, the solution collects aggregated data from a subset of nodes in the network without revealing which sensor nodes contributed to the aggregate. Identities of nodes participating in the data aggregation are not revealed even to other nodes that contribute to the data aggregation. 
As shown in Section~\ref{sec:analyses}, in passive attacks, the disclosure of details about sensor nodes or the data aggregation is contingent only on the likelihood of the query path traversing a sequence of nodes controlled by the attacker. 
Since the query path is randomly selected, the attacker can increase the likelihood of disclosing information only by increasing the number of controlled sensor nodes.

In addition, the solution is aggregating data without relying on aggregator nodes (apart from the sink node). Since aggregator nodes are aggregating data from multiple sensor nodes, they are an appealing target for attackers and a point of failure for the network.

\section{Research contribution}
\label{sec:research_contribution}

\subsection{The WSN model}
\label{sec:contrib}
Wireless sensor networks are commonly consisting of devices constrained in processing, bandwidth, storage, and energy resources. 
The literature adopted the term ad-hoc because it is often required that a WSN starts functioning by itself right after the deployment without a priori knowledge of the physical location. 
Nodes must wake up, identify neighbors, and set up a network without infrastructure. However, design constraints are application dependent and are based on the monitored environment~\cite{yick2008wireless}.

Throughout the manuscript, we consider a WSN as a wireless multi-hop network consisting of two types of nodes. The majority are nodes equipped with sensing technology, limited in computational capacity and memory as they are designed to be cheap. We refer to these nodes as sensor nodes. The other type of node is named sink node, which purpose is to gather data sensed by sensor nodes and act as a gateway to external systems. Moreover, the sink node has greater computational and storage capabilities than sensor nodes. Concerning the building monitoring scenario, we assume that an attacker cannot compromise sink nodes since sink nodes should be adequately secured and located in sections of the building with restricted access.
The network relies on a routing protocol for multi-hop wireless networks, the IP protocol, and a secure data link layer. The proposed privacy-preserving solution works over the TCP protocol. 

Nodes in the WSN are configured at deployment with a static IP address and a public-private key pair.
Sink nodes are configured with the following information of sensor nodes: IP address, public key, sensed physical quantities, and location of the building in which the sensor node is positioned (e.g., \textit{room237}). In building monitoring scenarios, new sensor nodes are rarely added to the network. Therefore, sensor node information can be added to sink nodes only by authorized personnel.

\subsection{The privacy-preserving communication protocol}
\label{sec:comm_protocol}

The network operates following an on-demand model: requests are issued to sink nodes from actors external to the network using a Domain-Specific Language (DSL)~\cite{fowler2010domain} for querying WSNs.
A sink node executes a request starting by translating a request into one or multiple query definitions as explained in Section~\ref{sec:request_processing}.
Queries are then constructed following the procedure described in Section~\ref{sec:query_construction}; each query consisting of a head and a body.

\begin{itemize}
    \item \textbf{Head:} an onion-like structure made of encryption layers, the head is of fixed size $L_{H}$ bytes. Each layer is intended to be decrypted by a sensor node in the query path. Layer decryption reveals the next-hop address (IP) or the next-hop address and a pair of symmetric encryption keys.
    \item \textbf{Body:} is consisting of the task $t$ specified in a general-purpose programming language and $w$ a fixed-size binary string used to transport task execution results back to the sink node that issued the query. The query body is encrypted using symmetric encryption and is of fixed size $L_{B}$ bytes.
\end{itemize}

The sink node in charge of the request will issue queries to the network. Queries will follow the query path encoded in the query head, query processing at sensor nodes is explained in Section~\ref{sec:query_processing}, and it branches based on the case that query head decryption reveals symmetric encryption keys. We refer to \textit{decoy nodes} as the sensor nodes in the query path that do not receive symmetric encryption keys and do not participate in the data aggregation. Moreover, we refer to \textit{target nodes} as sensor nodes in the query path that receive symmetric encryption keys; thus, they can decipher the query body and participate in the data aggregation.

All queries travel a path forming a circuit that ends at the sink node that issued the query.
Therefore, the sink node that issued a set of queries to accomplish a request will wait until it gets all queries' results.
Results are then merged to obtain the request result, which is forwarded to the external actor that triggered the request as explained in Section~\ref{sec:result_retrieval}. 
Throughout this section we use the notation in Table~\ref{tab:definitions}.

\begin{table}[t]
\scriptsize
\caption{Notation used in Section~\ref{sec:comm_protocol}.}
\label{tab:definitions}
\begin{tabular}{lp{7cm}@{}}
\toprule
\textit{\textbf{Symbol}}      & \textit{\textbf{Description}} \\ \midrule
\multicolumn{1}{l|}{$R$} &  a request specified as the operation $\phi$ and its target $\tau $                \\
\multicolumn{1}{l|}{$U$} &  set of all sensor nodes in the WSN   \\
\multicolumn{1}{l|}{$Q$} &  set of sensor nodes target of the request  \\
\multicolumn{1}{l|}{$P$} & set of query definitions $(S,K,e_{F},e_{L}) \in P$      \\
\multicolumn{1}{l|}{$t$} &   the task, computer code of size $L_{t}$ bytes              \\
\multicolumn{1}{l|}{$w$}  &        binary string of fixed size $L_{w}$ bytes \\ 
\multicolumn{1}{l|}{$\pi$} &set of recovery rules               \\
\multicolumn{1}{l|}{$n$} &      query path length                \\
\multicolumn{1}{l|}{$S$} &   list of sensor nodes in the query path                   \\
\multicolumn{1}{l|}{$s_{i}$} &    the $i-th$ sensor node in the list $S$               \\
\multicolumn{1}{l|}{$K$} &    list of pairs of symmetric encryption keys                  \\
\multicolumn{1}{l|}{$L_{H}$} &      query head size in bytes                \\
\multicolumn{1}{l|}{$B$} &      query body encrypted of fixed size $L_{B}$ bytes              \\
\multicolumn{1}{l|}{$OR_{i}$} &    query head encryption layer     \\
\multicolumn{1}{l|}{$\varepsilon(.)$} &          public-key encryption            \\
\multicolumn{1}{l|}{$X_{s{i}},Y_{s_{i}}$} & $X_{s{i}}$ private and $Y_{s_{i}}$ public key of the sensor node $s_{i}$                    \\
\multicolumn{1}{l|}{$E(.)$} &       symmetric key encryption               \\
\multicolumn{1}{l|}{$e$} &     symmetric encryption key                 \\
\multicolumn{1}{l|}{$p,\lambda$} &      padding                \\
\multicolumn{1}{l|}{$\Delta_{t},\Delta_{q}$}  &  time intervals in milliseconds  $\Delta_{t} < \Delta_{q}$                    \\ 
\bottomrule
\end{tabular}
\end{table}

\subsubsection{Request processing}
\label{sec:request_processing}


This section describes the operation of the sink node upon reception of a request.
A request $R$ is enclosing information about the operation and its target. We denote with $ \phi $ the operation detailed by the DSL code. 
Although the proposed solution allows conveying general-purpose computer code to sensor network nodes, the set of supported operations is bounded by the design of our protocol. Each query traverse several sensor nodes, and at each target node, the operation $\phi$ is processed, acquiring input from sensors equipped on the sensor node and from $w$ the binary string that carries results of $\phi$ processed on the previous target node. Therefore, the solution is supporting operations that produce a partial result while feeding on the input of sensor values and the previous partial result. Between the common supporting operations, we could list average, sum, max but also variance and standard deviation since in \cite{castelluccia2005efficient} was shown how to compute them as additive aggregation\footnote{In additive aggregation, a sensor node sums its sensed value with a received partial result, and forwards the sum to the next sensor node.}. Moreover, our solution allows to pose conditions on the data to be aggregated, like the exceeding of a threshold value or past sensed events. Conditions can be posed not solely on the physical quantity to retrieve but also on the status of other sensor technology equipped onto sensor nodes. 
The operation target $\tau$ specifies one or multiple locations of the WSN from which data will be aggregated. A possible request could consist of: e.g., $\phi =$  \texttt{IF(light==ON) THEN AVG(temperature)} and \texttt{$\tau = \langle$ room237, laboratory2 $  \rangle$}.

The operation $ \phi $, and the target $\tau$ are fed into the request translation component. We generalize the request translation component using the function: $f:(\phi , \tau ) \longrightarrow  t,P,\pi$.
The request translation component converts a request $R$ consisting of the operation $ \phi $, and its target $\tau$, into a task $t$ specified in a general-purpose programming language, a set $P$ consisting of one or multiple query definitions, and $\pi$ a set of recovery rules. 

First, the operation $\phi$ is used to generate the task $t$.
Then $\phi$ and $\tau$ are used to construct $Q \in U$, the set of sensor nodes that are targets of the request. $U$ is the set of all sensor nodes in the network.
The set $Q$ is constructed by selecting sensor nodes from the set $U$ that meet the location detailed in $\tau$ and sensed physical quantities required in $\phi$.
The set $P$ is then constructed with the algorithm \ref{alg:query_path_selection} repeated until $Q = \emptyset$. Algorithm \ref{alg:query_path_selection} will eventually empty the set $Q$, since every call of the function \textit{pickTarget()} removes a target from $Q$ and inserts it in the query path.
Each run of the algorithm will construct a query definition detailed by the tuple  $(S,K,e_{F},e_{L})$.

\begin{itemize}
    \item $S=\left \langle s_{1},...,s_{n}\right \rangle$ a list consisting of sensor nodes belonging to the set $U$. The list $S$ defines the query path.
    \item $K=\left \langle k_{1},...,k_{n}\right \rangle$ a list of elements $k_{i}$, where $k_{i}=(e_{a},e_{b})$ if $s_{i}$ is a target node, otherwise $s_{i}$ is a decoy node and $k_{i}= null$. ($k_{i}$ and $s_{i}$ the $i$-th elements of their respective list $K$ and $S$, and $(e_{a},e_{b})$ a pair of not equal symmetric encryption keys.) Moreover, encryption keys in the set $K$ are arranged as follows: if $s_{i}$ is a target node and $s_{j}$ is the next target node in the query path, then $k_{i}=(e_{a},e_{b})$ and $k_{j}=(e_{b},e_{c})$.
    \item $e_{F}$ the first symmetric key. If $s_{i}$ is the first target node in the query path $S$, then $k_{i}=(e_{F},e_{x})$.  
    \item $e_{L}$ the last symmetric key. If $s_{i}$ is the last target node in the query path $S$, then $k_{i}=(e_{y},e_{L})$.  
\end{itemize}

By our solution design, the query path length is a fixed network parameter; therefore, each query definition constructed using algorithm \ref{alg:query_path_selection} will include $n$ nodes in its path.
However, the query path not only includes nodes that contribute to the request (called target nodes), but also decoy nodes. Target nodes are sensor nodes in the query path that will receive symmetric encryption keys and will participate in the data aggregation process.
Therefore, $s_{i}$ is a target node if $k_{i}=(e_{a},e_{b})$ otherwise $s_{i}$ is a decoy node and $k_{i}= null$. ($k_{i}$ and $s_{i}$ the $i$-th elements of their respective list $K$ and $S$)
Algorithm \ref{alg:query_path_selection} iteratively constructs a query definition using two \texttt{while} loops.
The first loop will inserts $\left \lfloor n/2 \right \rfloor$ target nodes at random positions in the query path. Uncertainty is introduced to prevent queries from having a predictable disposition of target and decoy nodes. Since the last node in the query path can identify its function of being the node that will forward the query back to the sink node, by algorithm \ref{alg:query_path_selection} this node is always a decoy node.
The second loop will fill the query path with randomly chosen decoy nodes. 

Besides query path, algorithm \ref{alg:query_path_selection} also constructs $K$ the list of elements $k_{i}=(e_{a},e_{b})$, where $(e_{a},e_{b})$ is a pair of not equal symmetric encryption keys. Symmetric encryption keys are delivered in pairs to target nodes. A target node $s_{i}$ receiving symmetric keys $(e_{a},e_{b})$ will use the first key $e_{a}$ to decypher the query body, then it processes the query body, and it will use the second key $e_{b}$ to encrypt the query body. Therefore, symmetric encryption keys in the query path are arranged following the rule: if $s_{i}$ is a target node and $s_{j}$ is the next target node in the query path, then $k_{i}=(e_{a},e_{b})$ and $k_{j}=(e_{b},e_{c})$. Algorithm \ref{alg:query_path_selection} also outputs $e_{F}$ and $e_{L}$. $e_{F}$ is the first symmetric encryption key assigned to the first target node in the query path, this key is used by the sink node at query construction to apply the first encryption layer on the query body. $e_{L}$ is the second symmetric encryption key assigned to the last target node in the query path, this key is used by the sink node to decrypt the query body, and retrieve the query result. 
Additionally, the key $e_{L}$ also acts as the query identifier and is mapped into $\pi$ the set of recovery rules.

Since a request will inquire data from a large set of nodes, and each query can include at most $\left \lfloor n / 2  \right \rfloor$ target nodes in its path, the typical request will be accomplished by issuing multiple queries.
The set of recovery rules $\pi$ holds identifiers of queries issued to accomplish one request.
Moreover, $\pi$ must also include information about the operation $\phi$ since different aggregation functions require different procedures to aggregate partial results.

\begin{algorithm}[t]
\scriptsize
\begin{algorithmic}
\Require $U$ set of all WSN sensor nodes \newline \hspace*{2.9em} $Q$ set of sensor nodes target of the request \newline  \hspace*{3em} $n$ query path length 

\Ensure $S$ path of the query \newline  \hspace*{2.25em} $K$ list of encryption keys \newline  \hspace*{2em} $e_{F}$ first symmetric encryption key \newline   \hspace*{2em} $e_{L}$ last symmetric encryption key
 \item[]
    \Procedure{random}{}
    \State\Return a float from an uniform distribution bounded by (0,1) 
  \EndProcedure
  \Procedure{pickDecoy}{}
    \State Chose randomly $s \in U \setminus  (Q \cup B)$, add $s$ to $B$ 
    \State \Return $s$
  \EndProcedure
  \Procedure{pickTarget}{}
    \State Chose randomly $s \in Q$, remove $s$ from $Q$
    , add $s$ to $B$
    \State \Return $s$
   \EndProcedure
  \Procedure{generateSymKey}{}
    \State \Return a valid symmetric encryption key
   \EndProcedure
  \item[]

  \State \noindent \textbf{Compute:}
  \State $B = \emptyset$\;
  \State $S,K = <empty>$\;
  \State $k = null$\;
  \State $e_{F} = $\Call{generateSymKey}~\;
  \State $e_{L}$ = $e_{F}$\;
  \State $i,t = 1$\;
  \State $l= min(\left \| Q \right \|, \left \lfloor n/2 \right \rfloor)$\;
  \item[]

   \While{$i \leq l$}
   \State $t = \left \lceil \Call{random} ~ * (n-1) \right \rceil$\;
   
    \If{S[t] == null}
      \State $S[t] = \Call{pickTarget} ~ $\;
      \State $i++$\;
    \EndIf
    \State \textbf{end if}
    \EndWhile
    \State \textbf{end while}    
  \State $i = 1$\;
  
  \While{$i \leq n$}
    \If{S[i] != null}
     \State k = ($e_{L}$,\Call{generateSymKey}~)\;
     \State $e_{L}$ = k[2]\;
    
      \Else
    \State  S[i] = \Call{pickDecoy}~\;
    \State  k = null\;
    
    \EndIf
    \State \textbf{end if}
    \State $K[i] = k$\;
    \State $i++$\;
    \EndWhile
    \State \textbf{end while}

    \caption{Query path selection}
    \label{alg:query_path_selection}
    \end{algorithmic}
\end{algorithm}

After processing $ \phi $ and $\tau$, the request translation component produces $P$ -- a non empty set of tuples of cardinality $ \left \| P \right \| = \left \lceil \frac{\left \| Q \right \|}{\left \lfloor n / 2  \right \rfloor} \right \rceil$, a task $t$ specified in a general purpose programming language, and $\pi$ a set of recovery rules. For each tuple $(S,K,e_{F},e_{L}) \in P$ a query is constructed as explained in Section~\ref{sec:query_construction}.

\subsubsection{Query construction}
\label{sec:query_construction}

In this section, we present how the sink node that received the request $R$ converts a query definition detailed by the tuple $(S,K,e_{F},e_{L})$, and a task $t$ into a query consisting of the \textit{head} and the \textit{body}.
We will refer to $ \varepsilon ( \cdot )$ to denote the encryption operation using public-key cryptography, and $ E ( \cdot )$ to denote the encryption operation using symmetric cryptography. 

\paragraph{Head construction}

The query head construction starts from $OR_{n+1}$, the innermost encryption layer of the onion-like structure. This layer's purpose is to forward the query identifier back to the query issuer securely.
We refer to the query issuer as the sink node that constructs the query and dispatches it to the network.
The innermost onion layer is formed via encryption of the query identifier $e_{L}$ and the padding $p$ using the issuer's public key $Y_{sink}$.
The padding $p$ is introduced to maintain the head of fixed size if the query includes fewer than $\left \lfloor n / 2  \right \rfloor$ target nodes. 
The following equation describes how to compute the innermost onion layer.

\begin{equation*}
    OR_{n+1} = \varepsilon_{Y_{sink}}(e_{L},p)
\end{equation*}

Next, the sink node will compute the layer $OR_{n}$. This layer is closing the circuit, forwarding the query back to the sink node. The layer is committed to $s_{n}$ the last sensor node of the list $S$. The layer $OR_{n}$ is computed like the layer $OR_{i}$,  with the sole exception of including the sink node ip address $ip_{sink}$ as the next-hop address. Therefore, we omit explaining layer $OR_{n}$ construction, and we give layer $OR_{i}$ construction in the following lines.

The layer $OR_{i}$ addressed to the sensor node $s_{i} \in S$ ($i$ as index of the $i$-th element in lists $S,K$ and index of the $i$-th encryption layer of the query head) is computed in two distinct ways. \textit{A:} following equation \ref{eq:1} if $k_{i} = null$, therefore node $s_{i}$ is a decoy node. Layer $OR_{i}$ is computed via encryption of the next hop ip address $ip_{s_{i+1}}$, and previous onion layer $OR_{i+1}$ using the public key $Y_{s_{i}}$ belonging to the sensor node $s_{i}$. \textit{B:} equation \ref{eq:2} is applied if $k_{i} = (e_{a},e_{b})$, therefore node $s_{i}$ is a target node. Layer $OR_{i}$ is computed via encryption of the next hop ip address $ip_{s_{i+1}}$, the two symmetric encryption keys $e_{a}$ and $e_{b}$, and the previous onion layer $OR_{i+1}$ using the public key $Y_{s_{i}}$ belonging to the sensor node $s_{i}$.

\begin{equation}
    OR_{i} = \varepsilon_{Y_{s_{i}}}(ip_{s_{i+1}},OR_{i+1})\label{eq:1}
\end{equation}

\begin{equation}
    OR_{i} = \varepsilon_{Y_{s_{i}}}(ip_{s_{i+1}},e_{a},e_{b},OR_{i+1})\label{eq:2}
\end{equation}

The layer construction repeats until the formation of $OR_{1}$, the head's first encryption layer, which is always of size $L_{H}$ bytes.

\paragraph{Body construction}

The query body $B$ includes the task $t$ and a fixed size binary string. Since the query body must be of fixed size $L_{B}$ bytes, and the size of task $t$ can vary, additional padding $p$ of $L_{t} - size(t)$ bytes must be included into the query body. ($L_{t}$ the maximum allowed task size in bytes, and $size()$ the function that returns the number of bytes of the given argument) Then the query body is constructed by encrypting the binary string $w$, the task $t$, and padding $p$ using the symmetric encryption key $e_{F}$. Query body construction can be summarized using the following equation:

\begin{equation*}
    B = E_{e_{F}}(w,t,p)
\end{equation*}

Now the query is complete: $OR_{1}$ the query head and $B$ the query body.


\subsubsection{Query processing}
\label{sec:query_processing}

After the query construction process, the sink node sends the query to $s_{1} \in S$, the first node in the query path. When a sensor node $s_{i} \in S$ receives the query ($i$-th node in the list $S$), it performs the following steps: \textit{query decryption}, \textit{task execution}, and \textit{query forwarding}.

\paragraph{Query decryption}

The sensor node $s_{i}$ decrypts the query head $OR_{i}$ using its private key $X_{s_{i}}$. Query head decryption reveals the next hop IP address $ip_{s_{i+1}}$, the next onion layer $OR_{i+1}$, and if $s_{i}$ is a target node, head decryption also reveals the pair of symmetric encryption keys $(e_{a},e_{b})$.

\paragraph{Task execution}

If the sensor node $s_{i}$ received symmetric encryption keys from query head decryption, then $s_{i}$ is a target node and will perform the following steps. Otherwise, if $s_{i}$ is a decoy node, it will skip the following steps to perform the step \textit{query forwarding}. 

The sensor node $s_{i}$ decrypts $B$ (the query body) using the first symmetric key $e_{a}$ revealing: the data-carrying string $w$, the task $t$, and the padding $p$. The task $t$ gets executed in the task execution environment with $w$ given as an argument. A task is executed at most for $\Delta_{t}$ milliseconds otherwise, task execution is interrupted. Task execution returns $w^{'}$, a binary string holding task execution results. 
Then $s_{i}$ constructs $B^{'}$ the query body consisting of the data carrying string $w^{'}$, the task $t$, and the padding $p$ all encrypted using the second symmetric encryption key $e_{b}$. Therefore, $B^{'}$ is constructed as follows: $B^{'}=E_{e_{b}}(w^{'},t,p)$. Since the content of $B^{'}$ differs from $B$ only in the binary string, but the binary string $w^{'}$ is of the same size of $L_{w}$ bytes as $w$, then query body size is maintained uniform.

\paragraph{Query forwarding}

Query head is reassembled by applying the technique for onion size uniformity introduced in \cite{syverson1997anonymous}. The query head size is maintained fixed at $L_{H}$ bytes by adding $\lambda $ a padding of $size(OR_{i})-size(OR_{i+1})$ random bytes at the end of the onion layer $OR_{i+1}$. Therefore, the query head is now consisting of $OR_{i+1} + \lambda $.
After $\Delta_{q}$ milliseconds ($\Delta_{t} < \Delta_{q}$) from receiving the query, the sensor node will randomly choose $r$ a \texttt{float}, and will wait for other $r \times \Delta_{q}$ milliseconds before forwarding the query to the next hop. 
Adequate bounding values should be selected for the randomly chosen $r$, e.g. $0 \leq r \leq 4$.  
After waiting the required time, the query made of the head $OR_{i+1} + \lambda $ and the body $B^{'}$ (or $B$ if node $s_{i}$ is a decoy node) is forwarded to the next hop $s_{i+1}$ at the IP address $ip_{s_{i+1}}$.

\subsubsection{Result retrieval}
\label{sec:result_retrieval}

Each query sent from the sink node to accomplish the request $R$ will follow a path forming a circuit that ends at the sink node issuer of the query. 
The sink node decrypts the query head consisting of the onion layer $OR_{n+1}$ revealing $e_{L}$ the symmetric encryption key that acts as the query identifier. The data-carrying string $w^{''}$ is then obtained from the query body via decryption using the symmetric encryption key $e_{L}$. 
When the sink node gathers the feedback of all queries issued to accomplish $R$ it starts the recovery process of the request result.
Query results are merged following recovery rules $\pi$ to obtain the end result of the request $R$. The request result is then forwarded to the external actor that inquired the sink node.

\section{Privacy-preservation analysis}
\label{sec:analyses}

In this section, we examine the communication protocol to identify vulnerabilities concerning privacy preservation. 
Specifically, we consider passive attacks, where the attacker's goal is to obtain important information while remaining unnoticeable~\cite{gao2018mobile}, without altering the network traffic or interfere with the normal functioning of the WSN. 
Even though passive attacks do not directly cause damage at the WSN, they can facilitate other security breaches. Therefore, passive attacks are of particular concern in WSNs deployed for building monitoring.
To carry out this investigation we consider external and internal privacy~\cite{li2009privacy,bista2010privacy,de2009privacy,zhang2013preserving}. External privacy is threatened by actors outside of the network listening to the wireless communication, while internal privacy is threatened by trusted participating sensor nodes of the WSN. 

Throughout our analysis, we assume that sink nodes are secure, they do not collaborate with the attacker, they cannot be compromised, and their encryption keys are not disclosed to anyone. Moreover, we assume that the network implements security at the data link layer using encryption as most standards used in WSNs supports (e.g., the IEEE 802.11~\cite{ieee802}).

\subsection{External privacy}
\label{sec:external_privacy}

To examine external privacy, we visualize a foreign actor that is monitoring the network traffic by eavesdropping on wireless communications. We dub this actor the external adversary.
Eavesdropping is the intercepting and reading of messages by unintended receivers. Since the majority of wireless communications use the radio frequency spectrum to broadcast signals over the air, transmitted signals can be easily intercepted using adequate receiving equipment~\cite{wu2007survey}. However, security measures are commonly present already at the physical layer~\cite{shiu2011physical,wu2007survey}. Since delving deep into security measures of the physical layer is out of the scope of this paper, we assume that the external adversary is able to differentiate a transmission transferring query information from ordinary network management traffic by observing the transmission length.

Since the transmitted information is encrypted at the data link layer and query size is maintained uniform, the only disclosed detail of an intercepted query transmission is the effective transferring of the query from one node to another. The node receiver of the transmission is then processing the query (as explained in Section~\ref{sec:query_processing}) or re-transmitting it to another node (routing in multi-hop networks). 
However, processing the query introduces a delay, missing if it is just re-transmitted. Therefore nodes processing the query can be identified; nonetheless, the external adversary cannot differentiate decoy nodes from target nodes since the query processing time at both kinds of nodes depends upon a randomly chosen \texttt{float}. Obviously, the external adversary can draw a guess on whether the node processing a query is a target node. The success probability is $50\%$ since generally half the nodes in the query path are target nodes, but anyhow the external adversary cannot validate its guess.

We now consider an external adversary whose monitoring range covers the whole WSN; therefore, it can intercept the whole wireless traffic generated by the WSN. Hence, the adversary can track a query sourcing from the sink node and moving through the network by monitoring its transmissions. However, normally, the WSN traffic is not populated by only one query, and the randomized nature of the query path will make various queries mix at nodes on their route. Even though the adversary violates security measures of the physical layer, security at the data link layer is changing data by encryption before each transmission. 
Furthermore, query size is maintained uniform; therefore, it is hard for an external adversary to track how the query transit through the network since the adversary cannot distinguish between queries.

\begin{figure*}[th]
  \centering
  \subfloat[]{\includegraphics[width=0.4\textwidth]{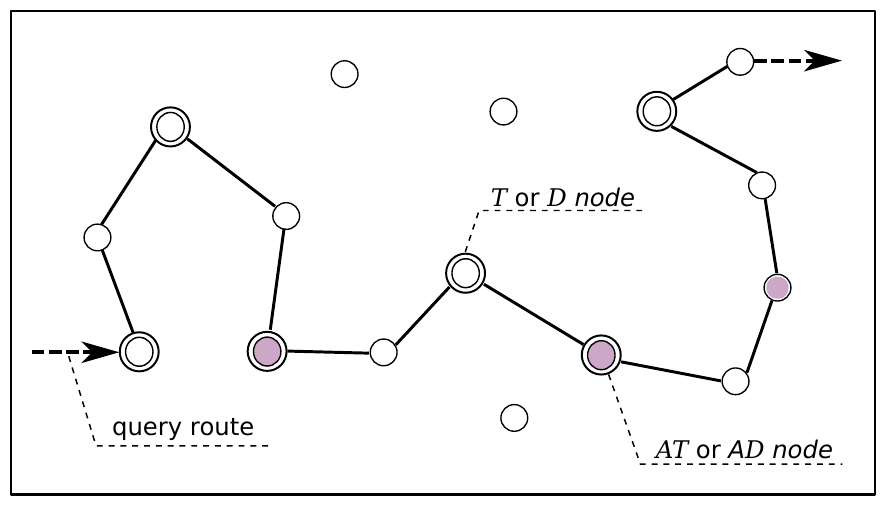}\label{fig:route1}}
  \hspace{40px}
  \subfloat[]{\includegraphics[width=0.4\textwidth]{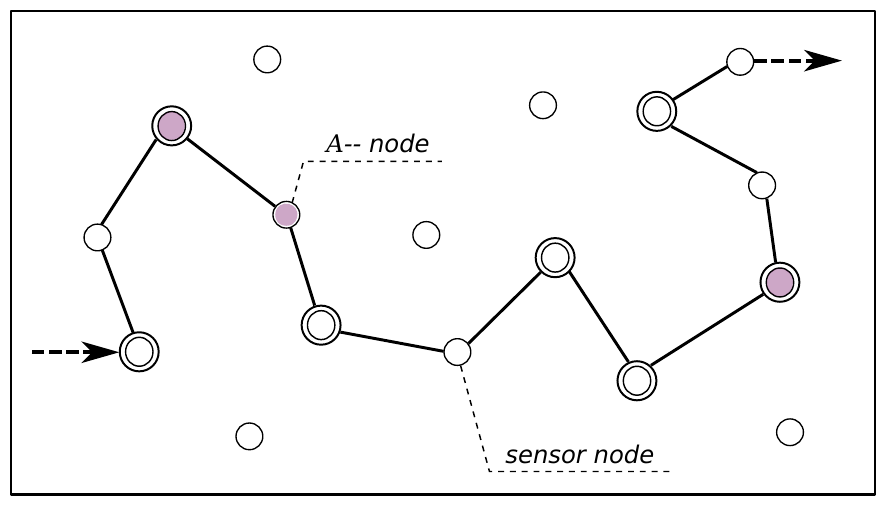}\label{fig:route2}}
  \caption{(a) Route of the query that leads through one $T|D$ node confined between two nodes owned by the adversary. (b) Route of the query that leads through two or more $T|D$ nodes confined between two nodes owned by the adversary. Node notation from table~\ref{tab:notations}.}
  \label{fig:internal_problems}
\end{figure*}

\subsection{Internal privacy}
\label{sec:internal_adversary}

An attacker that owns a subset of nodes of the WSN is commonly referred to as an internal adversary.
Nodes owned by an internal adversary are participating trusted nodes of the WSN owning cryptographic keys to decrypt messages addressed to them.
Although we assumed that the network implements security at the data link layer, only individual links are secured by such a solution, and nodes intermediate to routing paths can overhear messages passing through them.
Therefore, the internal adversary takes advantage of owned nodes to analyze traffic they receive and disclose information from un-compromised nodes of the WSN. 
In the following, we will analyze under which circumstances an internal adversary is able to gain insights over other nodes in the network and when the data privacy of a sensor node is disclosed.

\begin{table}[t]
\caption{Notations used in Section~\ref{sec:internal_adversary}, expressing the function of sensor nodes in relation to a query.}
\label{tab:notations}
\begin{tabular}{lp{7.25cm}@{}}
\toprule
\textit{Notation}       & \textit{Description} \\ 
\midrule
\multicolumn{1}{c|}{$D$} &  sensor node that imitates query execution (decoy node)   \\
\multicolumn{1}{c|}{$T$} &  sensor node target of the query    \\
\multicolumn{1}{c|}{$A$} &  sensor node owned by the adversary intermediate to the routing path of the query  \\
\multicolumn{1}{c|}{$AD$} & sensor node imitating query execution (decoy node) and owned by the adversary   \\
\multicolumn{1}{c|}{$AT$} & sensor node target of the query and owned by the adversary   \\
\end{tabular}
\end{table}

To formalize the analysis, we introduce the notation displayed in Table~\ref{tab:notations}. In the following, we explain the implications of a transiting query over sensor nodes. 

$T$ and $D$ are nodes that will process the query (as explained in Section~\ref{sec:query_processing}). On these nodes, the adversary is trying to gain information. Nodes $T$ are target nodes for the query, and after processing at $T$ nodes, the redirected query is entirely changed by encryption. On the other hand, after query processing at $D$ nodes, the redirected query has query head changed by encryption, but the query body remains unchanged.

$A$ are nodes owned by the adversary that are not processing the query, but they received the query due to routing needs in wireless multi-hop networks. Therefore, $A$ nodes receiving the query can observe the encrypted query head and query body. Moreover, the IP header reveals the IP address of the node that was previously processing the query and the IP of the next node that will process the query. 

$AD$ nodes are owned by the adversary and are processing the query. However, $AD$ nodes process the query as decoy nodes without decrypting the query body. Therefore, $AD$ nodes disclose only the node's IP address that was previously processing the query and the IP of the next node that will process the query. 

$AT$ nodes are owned by the adversary and are processing the query as target nodes. Therefore, nodes $AT$ can decipher the query head layer addressed to them, revealing the next-hop IP address and a pair of symmetric encryption keys. 
The internal adversary can use the first symmetric encryption key to decipher the query body and thus learn the task and the binary string that carries task execution results.
The internal adversary can examine the task code and disclose the aggregate to be computed; hence the adversary can identify the value carried in the binary string.
We say identify since the adversary can recognize that the value is a sum, an extreme, etc. Although the internal adversary can identify the value carried in the binary string, by owning a single node in the query path, the internal adversary cannot draw conclusions on the value or extract sensor node readings of other target nodes in the query path since it does not know which are target nodes and how many target nodes contributed to the aggregated value.

In the following, we consider the situation when a query is transiting through multiple nodes owned by the internal adversary. Hence the adversary can track sender and receiver IP enclosed in the IP header to partially reconstruct the query path and observe how the query body changes to gain information over other nodes in the query path. Note, we refer to the query path as the sequence of nodes on which the query is processed; on the other hand, we refer to the query route as the sequence of nodes that the query transits, including nodes in the query path and nodes that are forwarding the query due to routing needs in wireless multi-hop networks.
To conduct this investigation we examine two cases where the query route leads through two nodes owned by the adversary: 1) two or more $T|D$ nodes confined between two nodes owned by the adversary; (~$|$ used as logical OR) 2) one $T|D$ node confined between two nodes owned by the adversary. 

Fig.~\ref{fig:internal_problems} shows an example of the two cases.
We consider these two cases since if the query route leads through more than two nodes owned by the adversary, the instance can be generalized to multiple of the aforementioned cases. Moreover, in Section~\ref{sec:query_entry_exit}, we consider concerns of query exit and entry point.

\subsubsection{Route of the query that leads through two or more \texorpdfstring{$T|D$} nodes confined between two nodes owned by the adversary}

To examine this case, we consider a query route that leads through one node $A|AD|AT$, then through two or more $T|D$ nodes, and again through one $A|AD|AT$ node. We first look at the instance that nodes owned by the adversary are not both $AT$. 
Since the query is processing at two or more consecutive un-compromised nodes, IP information accompanying the query cannot be used to determine if both nodes owned by the adversary received the same query. 
Therefore, the adversary must rely solely on the 
query body to disclose meaningful information.
Indeed, if nodes in the query path arranged between the two nodes owned by the adversary are all $D$, the inner encryption layer of the query body will remain unchanged. 
Therefore, the adversary can identify that both owned nodes received the same query and that all nodes processing the query between the owned nodes are $D$ nodes.
However, the internal adversary is not able to recognize the number of nodes in the query path between the two owned nodes since query processing time depends upon a randomly chosen \texttt{float}. On the other hand, if any node arranged between the two nodes owned by the adversary is a $T$ node, then the query body is also changed by encryption, and the internal adversary cannot determine if both owned nodes are executing the same query.

We now consider that both nodes owned by the adversary are $AT$ nodes. Then the adversary can decipher the query body at both owned nodes, and it should be possible to compare tasks and binary strings to recognize if both nodes received the same query. However, by our solution design, multiple queries carry the same task, making it hard to exactly recognize if both nodes received the same query.
Even if we consider the case an adversary is able to tell that both owned nodes received the same query, the adversary cannot draw conclusions on nodes processing the query between the two owned nodes since the adversary cannot identify which of them are $T$ nodes. 
Even though the adversary can recognize the value change of the binary string, he cannot know how many $T$ nodes contributed to the value change.

\subsubsection{Route of the query that leads through one \texorpdfstring{$T|D$} node confined between two nodes owned by the adversary}

This disposition of nodes can be identified by the internal adversary as a transitive dependency of sender and receiver IP addresses from the IP packet header (e.g. $IP_{owned1} \rightarrow IP_{i} , IP_{i} \rightarrow IP_{owned2}$).
Therefore, we assume that given the transitive relation of IP addresses, the adversary deduces that the two owned nodes are processing the same query, even if the query was changed by encryption and the randomly chosen query processing time does not ensure that it is the same query. 
Another query might be routed through  $IP_{i} \rightarrow IP_{owned2}$ tricking the internal adversary of detecting the relation $IP_{owned1} \rightarrow IP_{i} , IP_{i} \rightarrow IP_{owned2}$.

Regardless of the aforementioned possibility of mixing queries, if the adversary identifies this particular disposition of nodes, he can examine the query received at the two owned nodes to gain insights over the node between them.
We distinguish the following three cases where the adversary discloses different insights over the un-compromised node between the two owned nodes: 

\begin{enumerate}
    \item \textit{Both nodes owned by the adversary are $A|AD$.} The adversary can compare the encrypted query body at both owned nodes to disclose if the un-compromised node is a $T$ or $D$ node for the received query. 
    
    \item \textit{One node owned by the adversary is $AT$.} The adversary can recognize if the un-compromised node is a $T$ or $D$ node for the received query.
    Furthermore, if the un-compromised node is a $T$ node, the adversary can observe the task code to disclose the sensor technology equipped on the node.
    
    \item \textit{Both nodes owned by the adversary are $AT$.} In this case, the adversary can gain insights summarized in the previous points, but additionally, if the un-compromised node is a $T$ node, the adversary can compare the binary string state at the two owned nodes to threaten the data privacy of the un-compromised node. 
\end{enumerate}

\subsubsection{Query entry \& exit point}
\label{sec:query_entry_exit}

Besides the two mentioned cases, we also analyze concerns related to the query entry and exit point. We refer to the query entry point as the first sensor node in the query path and the query exit point as the last node in the query path since they both communicate with the sink node making them recognizable as such.

By our solution design, the first node in the query path can be a $T$ node. Thus if the first and the second node in the query path are respectively $T$ and $AT$, the adversary can disclose data privacy of the first node in the query path, however only if the adversary can identify that he owns the second node in the query path. 
Since the position of nodes in the query path can be identified only by tracking the query route from the source, this vulnerability can be exploited only if the adversary owns an $A$ node in the query route from the sink node to the first node in the query path.

The query exit point or the last node in the query path is always a $D$ node. We included this design choice since the last node in the query path can identify its position in the query from the next-hop IP address, which is the address of the sink node.
If $AT$ is the last node in the query path able to decipher the query body, the adversary could potentially infer over the number of $T$ nodes that contributed to the aggregated result, since a query generally includes $\left \lfloor n / 2  \right \rfloor$ $T$ nodes in the query path. However, the adversary can identify that he owns the last sensor node deciphering the query body only by tracking the query path to the sink node.

\section{Results}
\label{sec:results}

This section presents results obtained from a simulation using the ns-3 simulator \cite{henderson2008network} of the privacy preserving communication protocol presented in Section~\ref{sec:comm_protocol}. The aim of this investigation is:
\begin{enumerate}
    \item Assess the scalability of the concept by examining how querying the WSN is affected by the following independent variables: query path length (number of nodes on which the query will be processed), and network size (number of nodes in the WSN), and WSN topology. 
    \item  Determine if the network topology significantly affects querying. 
\end{enumerate}

To conduct this investigation, we meter the Query-Time-To-Return (QTTR). We define the QTTR as the elapsed time between the issuing of the query from the sink node and the return of the issued query to the issuer node. Therefore, the QTTR includes the processing time of the query at sensor nodes in the query path. 
Although in the concept presented in Section~\ref{sec:query_processing} the query processing time depends upon a randomly chosen \texttt{float}, our simulation does not implement this feature since it introduces delays that are not dependent on the network. Therefore, in the simulation, a query processed at a sensor node is forwarded to the next-hop node right after query processing ends. 

\begin{figure*}[th]
    \centering
    \includegraphics[scale=1]{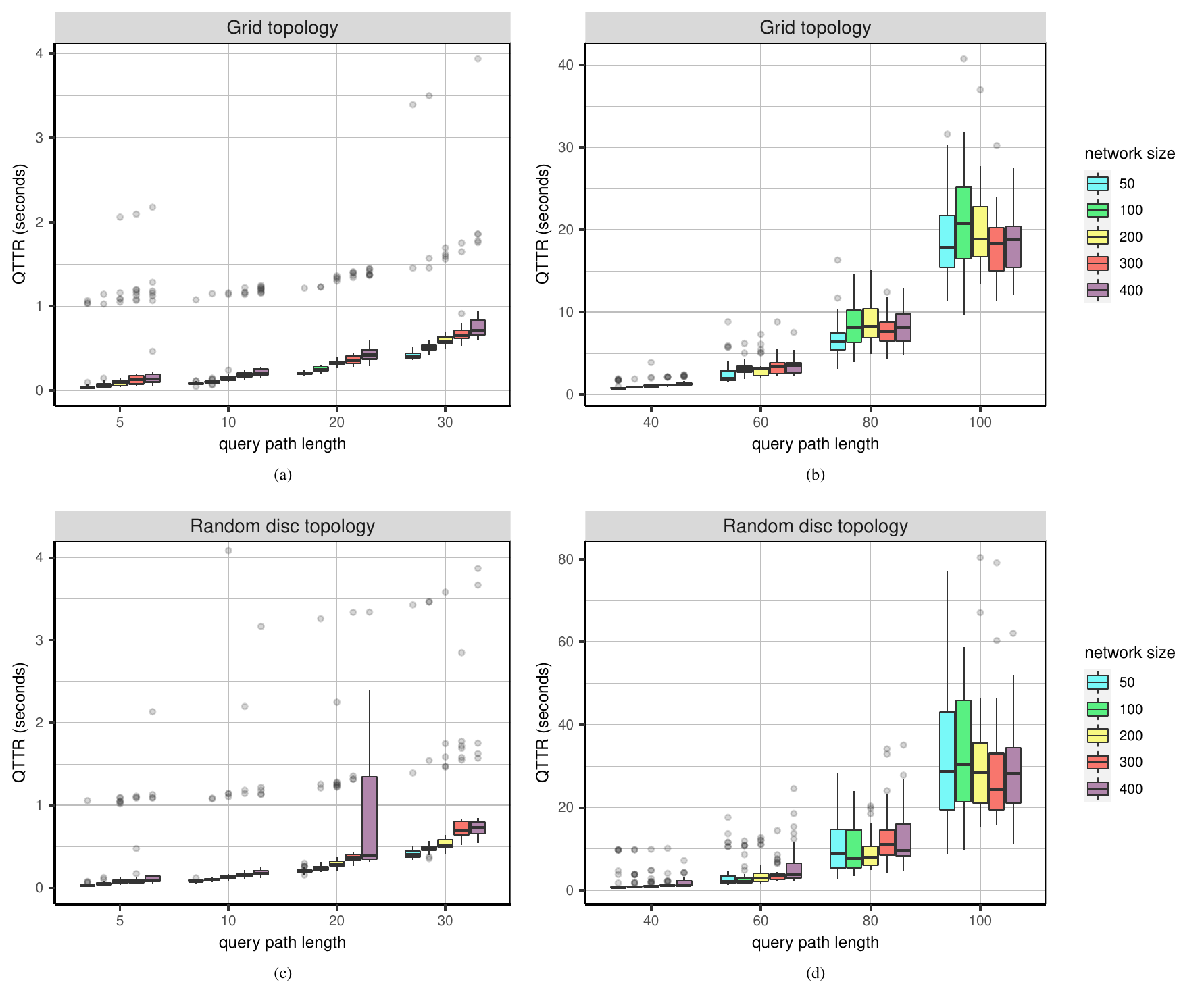}
  \caption{Charts show the box-plot representation of the QTTR at varying of the query path length and network size. Data from the experiment 1 in Section~\ref{sec:experiment1}.}
  \label{fig:overview}
\end{figure*}

\subsection{Experimental setup}

In our simulation, the WSN consists of one sink node and $s$ sensor nodes. We consider two network topologies: the \textit{grid topology} (GT) and the \textit{random disc topology} (RDT). In the former, sensor nodes are deployed according to a grid structure; each sensor node is equidistant from the closest sensor nodes in cardinal directions. We set the distance between sensor nodes to $a=60$ meters so that a sensor node is in the communication range of at most eight sensor nodes. In the latter, sensor nodes are randomly deployed on a disc-shaped plane of radius $r_{p}$. The radius $r_{p}$ is obtained from $r_{p}=\sqrt{A / \pi}$, $A$ being the sum of circular area's covered by $s$ sensor nodes at radius $r_{s}=35$ meters. Therefore, the average sensor node density of the network is maintained fixed at diverse $s$. Since in RDT sensor nodes are casually deployed on the target area, some sensor nodes could form independent network segments not connected to the network segment of which the sink node is a member. Therefore some sensor nodes might never be queried. 
In both network topologies, the sink node is deployed in the center of the WSN. 
We choosed values of parameters $a=60m$ and $r_{s}=35m$, since networks of different topologies will have a near equal average node density: GT: $\frac{1}{60^{2}}=0,277*10^{-3} nodes/m^{2}$, and RDT: $\frac{1}{35^{2}\pi }=0,26*10^{-3} nodes/m^{2}$.

The wireless communication in the simulated network adheres to the IEEE 802.11a standard for local wireless networks, operating at the data rate of 12Mbps. The maximum transmission unit is set to the ns-3 default 536 bytes. Each node in the WSN has installed the IP stack, and messages are transmitted over the TCP protocol. 

However, the TCP was designed to function over low-error wired networks where the packet loss is usually the outcome of a network congestion~\cite{chandran2001feedback}. Several studies are suggesting that the use of TCP in wireless multi-hop networks results in low throughput since packet loss due to transmission error and route discovery is handled using congestion control~\cite{al2005survey,liu2001atcp,chandran2001feedback,gomez2018tcp}.
Route discovery is performed by the routing protocol when searching for a route from sender to receiver. 
It is possible that discovering a route may take more time than the TCP retransmission timeout (RTO)~\cite{liu2001atcp}.
The RTO is an internal timer of the TCP used to determine when a segment needs to be retransmitted. If the RTO elapses before receiving the acknowledgment of segment delivery, the segment is retransmitted, the RTO is increased using exponential backoff, and the TCP is adjusted for congestion. The minimum RTO value in the simulation is set to the default, 1 second. 
To avoid complications due to route discovery, we decided to use the Optimized Link State Routing Protocol (OLSR)~\cite{clausen2003optimized}, a proactive routing protocol so that routes are immediately available when needed. In proactive routing protocols, routes between each pair of nodes are determined at the network start-up and maintained with periodic updates~\cite{abolhasan2004review}.

To encrypt query head layers we opted for Elliptic Curve Cryptography (ECC)~\cite{miller1985use} against Rivest-Shamir-Adleman (RSA)~\cite{rivest1978method} based cryptosystems. Several evidence ~\cite{lara2018elliptic,microchip} are suggesting the adequacy of ECC cryptosystems for resource constrained devices; mainly due to the smaller key size compared to RSA. Based on the guidelines defined by the National Institute of Standards and Technology (NIST)~\cite{barkernist}, the cryptosystem security is acceptable until 2030 and beyond if it offers a security strength of at least 128b. This security level is achieved by the RSA method at a key length of 3072b and by ECC methods at a key length of 256b. 

In our simulation, cryptographic operations are done using the Libsodium library~\cite{NaCi}. The public key cipher used to encrypt query head layers is based on Curve25519~\cite{bernstein2006curve25519} and has a key length of 256b. In the used Libsodium implementation named \textit{Sealed box}, the encryption operation will output a ciphertext that is 48bytes larger than the plaintext due to the shared secret.

The simulation works in two phases; in the first phase, all sensor nodes in the network send their public key to the sink node. The first phase aims to determine which sensor nodes are in the same network segment as the sink node. In the second phase, a set of queries is issued to the network, and data about queries is collected.
Queries are constructed from the sink node by randomly selecting $n$ nodes to include in the query path, $n$ being the query path length. The query path gets encoded in the query head, consisting of the onion-like structure made of encryption layers, each layer holding the next-hop IP address and the inner encryption layer. The query body consists of the binary string used to carry the aggregated result.

Queries are issued from the sink node sequentially; after a query returns back to the sink node, the following query is issued. 
A node in the query path that receives the query will decrypt the query head outer encryption layer to reveal the next-hop IP address and the inner encryption layer.
The received query head is replaced with the inner encryption layer, and the same number of bytes removed from layer decryption are included as padding to maintain uniform query size. After layer decryption, the node will add its sensed value to the binary string value and forward the modified query to the next-hop node. If the query does not reach the next-hop node in 30-seconds, the query is aborted, and a new query is issued from the sink node.

\subsection{Experiment 1 -- Scalability}
\label{sec:experiment1}

\begin{figure}[t]
    \centering
    \includegraphics[width=0.9\columnwidth]{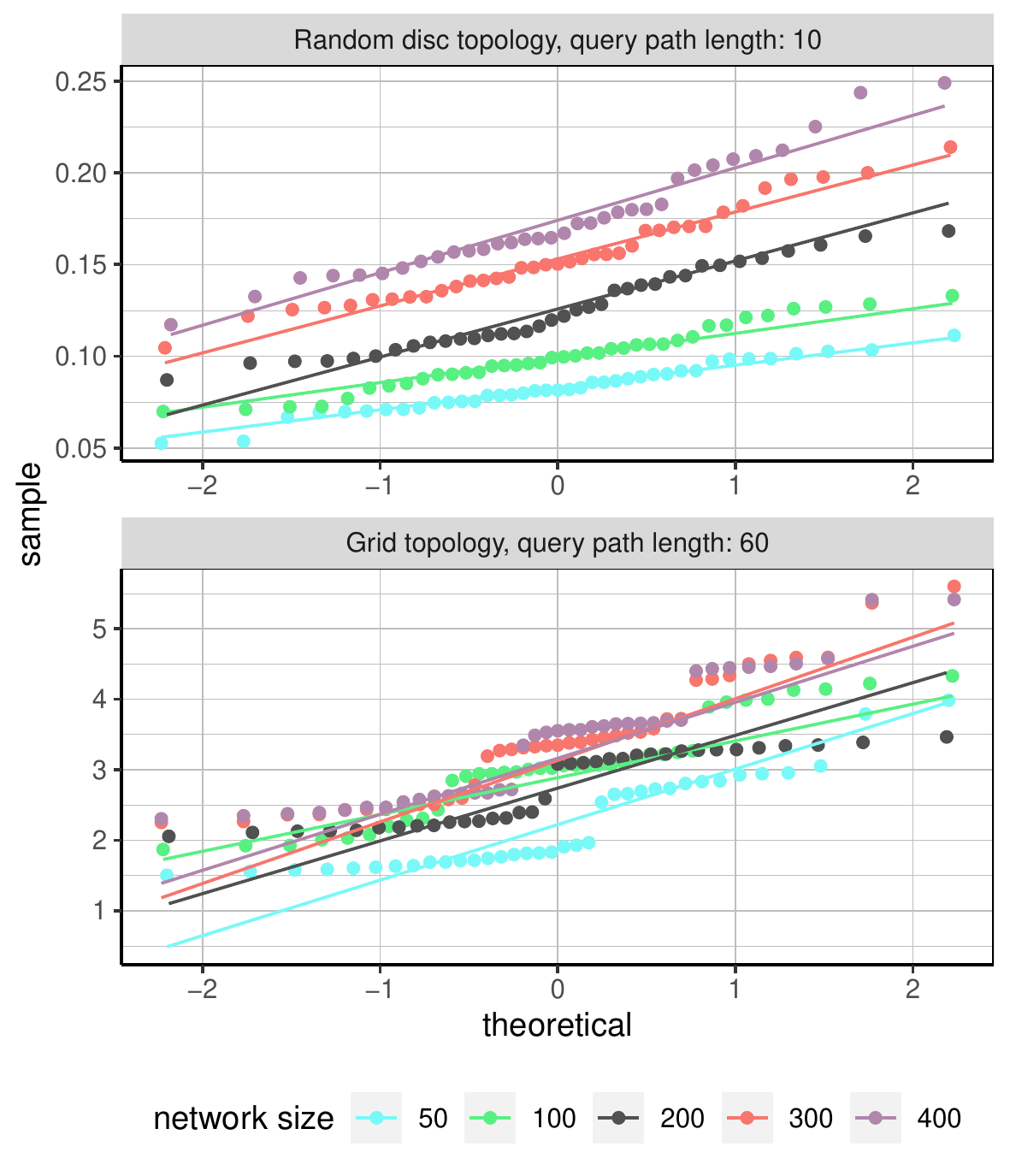}
    \caption{Quantile-quantile charts of QTTR per different network size. Data from experiment 1 in Section~\ref{sec:experiment1}.}
    \label{fig:qq-normality}
\end{figure}

\begin{figure}[t]
    \centering
    \includegraphics[width=\columnwidth]{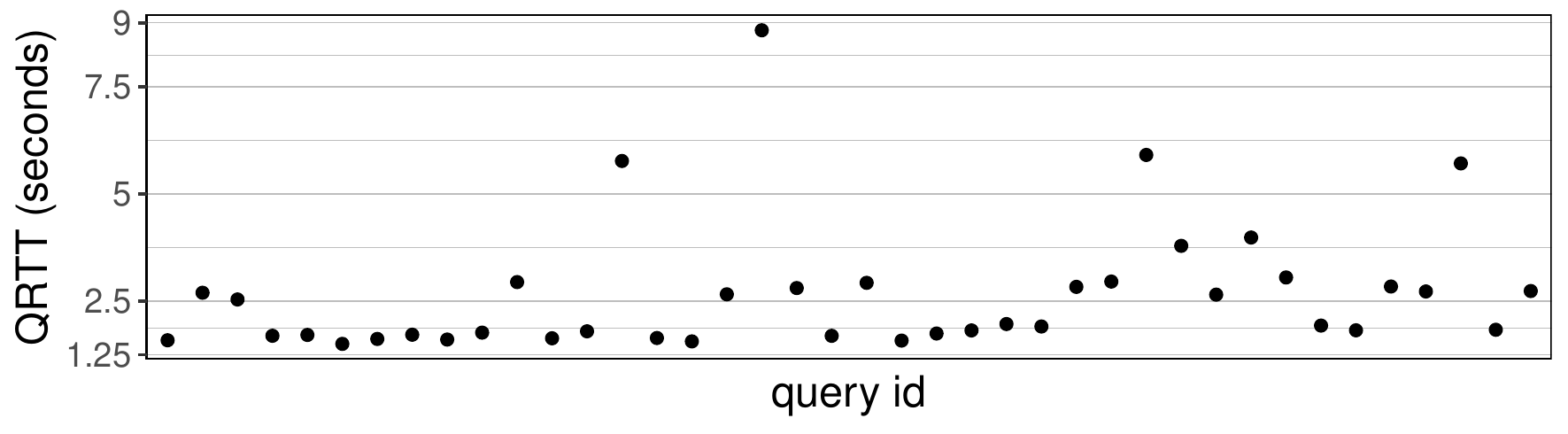}
    \caption{Scatterplot of the QTTR at independent variables: GT, network size of 50 nodes, and query path length of 60 nodes. Data from experiment 1 in Section~\ref{sec:experiment1}.}
    \label{fig:scatter}
\end{figure}

A set of simulations was run for both GT and RDT at $s=\left \{ 50,100,200,300,400 \right \}$. Each run executing 40 queries for each value of $n=\left \{ 5,10,20,30,40,60,80,100 \right \}$, the query path length. The obtained data is presented in Fig.~\ref{fig:overview}, and summary statistics are given in Table~\ref{tab:summary_grid} and~\ref{tab:summary_disc}. From Fig.~\ref{fig:qq-normality} can be seen that the QTTR does not follow a normal distribution at fixed query path length and network size.
In the scatterplot presented in Fig.~\ref{fig:scatter} we can see that at a query path length of 60 nodes in a network of 50 nodes based on GT deployment, the QTTR is mainly distributed in the 1.25-2.5 seconds interval. Other measures that do not fit in the 1.25-2.5 interval are likely a consequence of packet loss and the consequent TCP retransmission. 
Furthermore, we believe that the TCP-RTO is the reason behind the non-normal distribution of the QTTR. Measures affected by the TCP retransmission were not removed from the dataset during analyses.

To establish whether network size affects the QTTR, we performed the Kruskal Wallis test~\cite{hollander1973wolfe}. The test results presented in Table~\ref{tab:kruskal_wallis} show that at both network topologies at the query path length $<$ 100, there is a significant difference between mean ranks at diverse network sizes. Furthermore from charts in Fig.~\ref{fig:overview} it is possible to notice that the relation between network size and QTTR is not exponential but is very much like a linear relation. 

\begin{figure*}[th]
  \centering
  \subfloat[Grid topology]{\includegraphics[width=\textwidth]{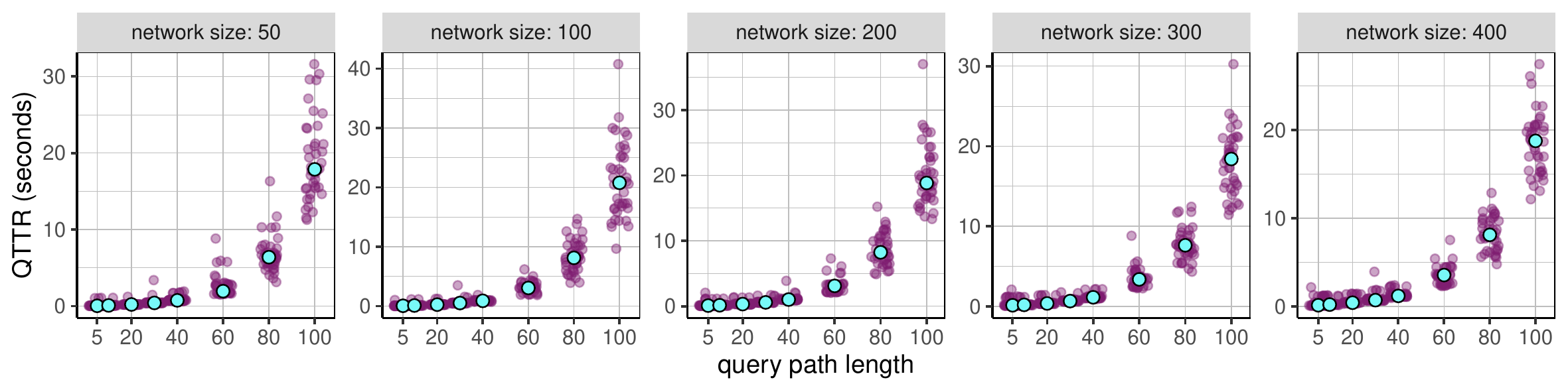}\label{fig:qlenGrid}}
  \hspace{40px}
  \subfloat[Random disc topology]{\includegraphics[width=\textwidth]{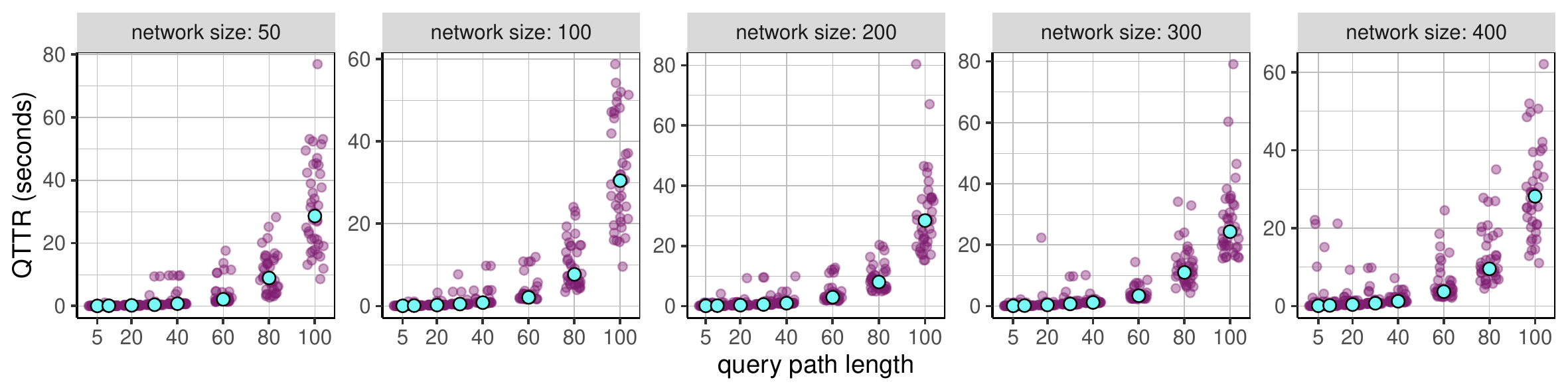}\label{fig:qlenDIsc}}
  \caption{Charts of QTTR in relation to the query path length. Blue dots are the medians. Data from experiment 1 in Section~\ref{sec:experiment1}.}
  \label{fig:qlen}
\end{figure*}

From the charts in Fig.~\ref{fig:qlen} it is possible to notice that the relation between the query path length and the QTTR is not linear. 
Probably, the non-linear relation is a cause of the query size increase due to a longer query path and because the query size is maintained uniform throughout the whole query path.

\subsection{Experiment 2 -- Network topology}
\label{sec:experiment2}

The second set of simulations was run to establish whether the network topology does affect the QTTR. We ran 30 simulations for both GT and RDT, each simulation having a network size of $s=200$ nodes and executing 40 queries of query path length $n=40$. 
Summary statistics of the collected dataset are presented in Table~\ref{tab:summary_topologyTest}. The Fig.~\ref{fig:topology_distributions} shows a density chart of the distribution of QTTR at both GT and RDT. 

However, applying the RDT results in networks with a diverse arrangement of nodes at each simulation run; therefore, there is a dependency between observations taken from a simulation run.
To conduct a statistical analysis we look at the article~\cite{galbraith2010study} which compares several approaches for clustered data analysis. 
Since the data is not normally distributed, clusters have the same number of observations, and the hypothesis of homogeneity of variance was rejected. 
We opted to reduce clusters to single observations using a summary statistic and then apply a statistical test. 
More specifically, since the data is not normally distributed, we summarized clusters using the median. 
Therefore, the output of each simulation run was reduced to a single value, the median of 40 measures of QTTR.
Reduced clusters were analyzed using the Brunner-Munzel test~\cite{brunner2000nonparametric} since cluster medians are not normally distributed, and homogeneity of variance was not confirmed. 
The test was not significant (Brunner-Munzel Test Statistic = $-1.3448$, $df = 36.989$, $p-value = 0.1869$, $\hat{p}" = 0.394$), test results are suggesting that the QTTR in networks with deployment GT is not significantly different than in networks with deployment RDT.

\begin{table}[t]
\centering
\caption{Kruskal Wallis test~\cite{hollander1973wolfe} results to determine if the mean ranks of QTTR per diverse network sizes are significantly different. The test was performed for both GT and RDT at each value of $n$. QTTR the measurement variable, and network size the nominal variable. Data from the experiment 1 in Section~\ref{sec:experiment1}}
\label{tab:kruskal_wallis}
\begin{tabular}{r|p{2cm}|p{2cm}}
\toprule
\multicolumn{1}{l}{}                       & \multicolumn{2}{c}{\textbf{network topology}}              \\ 
\cmidrule(l){2-3} 
\multicolumn{1}{c|}{\textbf{query path length}} & \multicolumn{1}{>{\centering\arraybackslash}p{2cm}|}{GT}                 & \multicolumn{1}{>{\centering\arraybackslash}p{2cm}}{RDT}          \\ 
\midrule
5                                          &          \multicolumn{1}{c|}{H=88.44,***}        &          \multicolumn{1}{c}{H=103.52,***}                           \\
~10                                        &         \multicolumn{1}{c|}{H=144.54,***}           &         \multicolumn{1}{c}{H=134.16,***}                              \\
20                                         &        \multicolumn{1}{c|}{H=144.26,***}              &         \multicolumn{1}{c}{H=134.37,***}                              \\
~30                                        &         \multicolumn{1}{c|}{H=127.26,***}             &        \multicolumn{1}{c}{H=104.91,***}                               \\
40                                         &          \multicolumn{1}{c|}{H=106.73,***}            &         \multicolumn{1}{c}{H=69.90,***}                              \\
60                                         &          \multicolumn{1}{c|}{H=31.45,***}            &         \multicolumn{1}{c}{H=32.87,***}                              \\
80                                         &           \multicolumn{1}{c|}{H=19.58,***}           &          \multicolumn{1}{c}{H=12.75,*}                             \\
100                                        &           \multicolumn{1}{c|}{H=7.01}           &           \multicolumn{1}{c}{H=4.13}                            \\ 
\midrule
\multicolumn{3}{l}{{\scriptsize significant: * at p \textless 0.05; ** at p \textless 0.01;*** at p \textless 0.001; \hspace{6px} 4 d.f.}}    \\
\bottomrule
\end{tabular}
\end{table}

\section{Discussion}
\label{sec:discussion}

\subsection{Discussion of the privacy-preservation analyses}

In Section~\ref{sec:external_privacy} we made clear that the solution withstands traffic analysis since queries follow all the same circuit-like patterns while paths are randomized, transmitted queries are indistinguishable by encryption and uniform query size, and queries are mixing while transiting the WSN. Furthermore, external actors observing the wireless communication cannot disclose the nodes' target of the query since query forwarding time is decoupled from query execution. 

However, we must point out concerns related to unbalanced querying of regions of the network. 
Querying a region of the network will involve $x$ nodes in query processing. 
$x/2$ nodes from the region of interest (target nodes) and $x/2$ nodes from random network locations (decoy nodes). Therefore if in a short time span a region of the network is queried more times than other regions, nodes located in the region of interest will be more involved in query processing than nodes in other regions of the network.
Indeed, an attacker monitoring the network traffic can rely on statistical analysis to identify the region of interest by detecting the increase in nodes performing query processing. Therefore, the querying of regions should be adequately balanced to prevent disclosures of regions of interest.

In Section~\ref{sec:internal_adversary} we showed that the proposed protocol preserves query privacy by constraining information of the query path. Nodes receiving the query that are able to decipher the query body can learn about the task. However, without knowledge of the identities of other nodes involved in the data collection, it is not possible to infer information other than those revealed by the task code. 

Analysis suggests that an attacker owning a portion of nodes of the network could possibly disclose insights about non-compromised sensor nodes and even threaten data privacy in the WSN.
However, sensor nodes forming the query path are randomly selected.
Therefore the odds of such disclosures are only related to the number of nodes owned by the adversary.

One solution to this problem is to rely on security control to add nodes to the network, such that an attacker can own a sensor node in the network only by taking control over a sensor node already associated to the WSN.
However, taking control over a sensor node requires physical access to the node~\cite{butun2019security}.

Furthermore, to mitigate complications described in Section~\ref{sec:query_entry_exit} occurring at query entry and exit point, the protocol can be extended to allow query processing at sink nodes and by setting an initial value to the data-carrying binary string accompanying the query.

\begin{table}[ht]
\centering

\caption{Summary statistics of the dataset collected in experiment 2, Section~\ref{sec:experiment2}. The second division present statistics on the number of nodes included in the same network segment as the sink node (nodes reachable by the query). The third division present statistics on the number of nodes in the one-hop neighborhood of nodes in the network.}
\label{tab:summary_topologyTest}
\begin{tabular}{r|p{1.2cm}|p{1.2cm}}
\toprule
\multicolumn{1}{l}{} & \multicolumn{2}{c}{\textbf{network topology}} \\ 
\cmidrule(l){2-3} 
\multicolumn{1}{c|}{\textbf{statistic}} & \multicolumn{1}{>{\centering\arraybackslash}p{1.2cm}|}{GT} & \multicolumn{1}{>{\centering\arraybackslash}p{1.2cm}}{RDT} \\ 
\midrule
\textbf{min} QTTR (seconds) & \multicolumn{1}{c|}{0.822} & \multicolumn{1}{c}{0.746} \\
\textbf{avg} QTTR (seconds) & \multicolumn{1}{c|}{1.509} & \multicolumn{1}{c}{2.053} \\
\textbf{max} QTTR (seconds) & \multicolumn{1}{c|}{6.694} & \multicolumn{1}{c}{15.224} \\
\textbf{std} QTTR (seconds) & \multicolumn{1}{c|}{0.832} & \multicolumn{1}{c}{2.070} \\
\textbf{q25} QTTR (seconds) & \multicolumn{1}{c|}{1.014} & \multicolumn{1}{c}{1.018} \\
\textbf{median} QTTR (seconds) & \multicolumn{1}{c|}{1.097} & \multicolumn{1}{c}{1.125} \\
\textbf{q75} QTTR (seconds) & \multicolumn{1}{c|}{2.002} & \multicolumn{1}{c}{2.096} \\
\midrule
\textbf{q25} \textbf{\%}~reachable nodes & \multicolumn{1}{c|}{100} & \multicolumn{1}{c}{88.5} \\ 
\textbf{median} \textbf{\%}~reachable nodes & \multicolumn{1}{c|}{100} & \multicolumn{1}{c}{92} \\ 
\textbf{q75} \textbf{\%}~reachable nodes & \multicolumn{1}{c|}{100} & \multicolumn{1}{c}{94.5} \\ 
\midrule
\textbf{q25} \#one-hop neighborhood & \multicolumn{1}{c|}{5} & \multicolumn{1}{c}{4} \\ 
\textbf{median} \#one-hop neighborhood & \multicolumn{1}{c|}{8} & \multicolumn{1}{c}{6} \\ 
\textbf{q75} \#one-hop neighborhood & \multicolumn{1}{c|}{8} & \multicolumn{1}{c}{12} \\ 

\bottomrule
\end{tabular}
\end{table}

\begin{figure}[ht]
    \centering
    \includegraphics[width=\columnwidth]{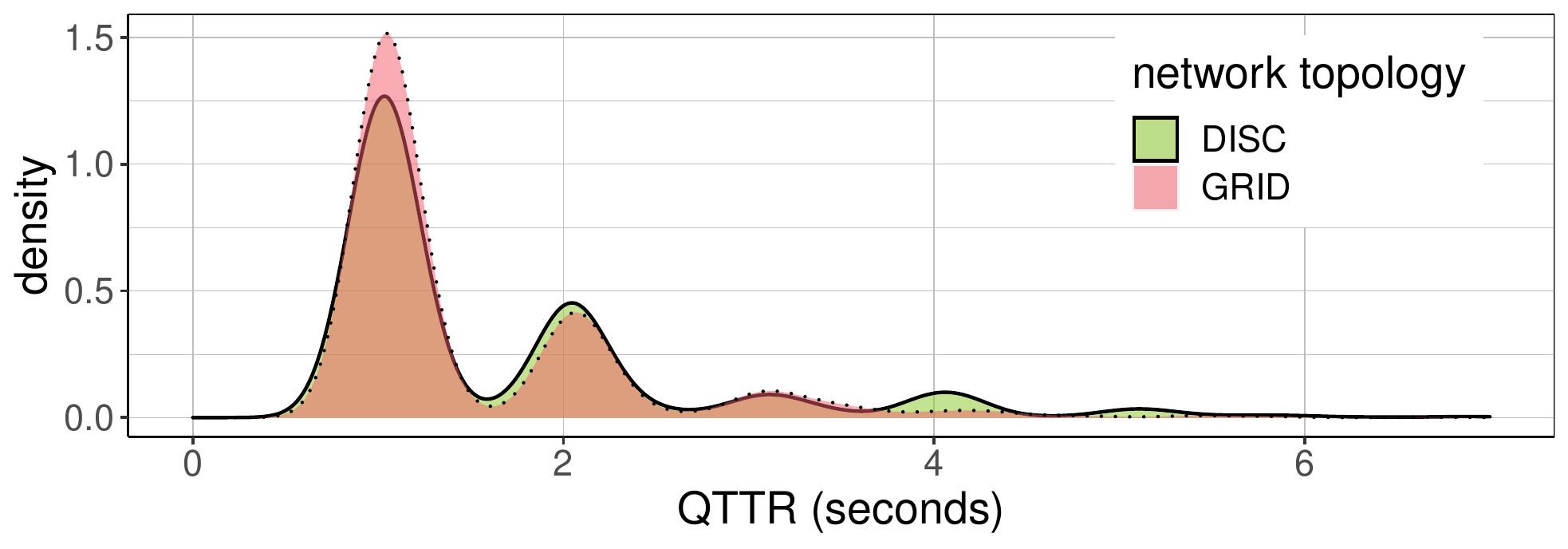}
    \caption{Density chart of the distribution of QTTR at network deployed according to GT and RDT. Data from experiment 2 in Section~\ref{sec:experiment2}.}
    \label{fig:topology_distributions}
\end{figure}

\subsection{Discussion of simulation results}

Results obtained from experiment 1 provide an overview of the scalability of the solution. Interestingly we noticed a linear relation between the network size and the QTTR. Therefore the data reported in this paper suggests that the protocol could be implemented in large WSNs. However, we suppose that a large WSN (more than 200 nodes) includes several sink nodes. Multiple sink nodes in the network allow a better load balancing, and the network keeps functioning even if a sink node ends operating. Additionally, in large WSNs with multiple sink nodes, it could be practical to constrain the network region that a sink node can query. This will diminish the QTTR as seen in Fig.~\ref{fig:overview} and will reduce the overall communication traffic due to the shortest routing paths of the query.

However, from Fig.~\ref{fig:overview}, and Table~\ref{tab:summary_grid} and \ref{tab:summary_disc} it is apparent that the QTTR is predominantly affected by the query path length. In our simulation, we opted for a public-key cryptosystem that integrates a key encapsulation mechanism with a data encapsulation mechanism~\cite{abe2008tag}. Consequently, each query head encryption layer must include an encapsulated secret; hence the query head size is largely affected by the key length of the selected public-key cryptosystem. In the simulation, each encryption layer increases the query head size by 48 bytes due to the shared secret and 4 bytes due to the IP address. Therefore the query head size is linearly increasing in relation to the query path length. However, the query size is maintained uniform by adding padding throughout the whole query path. We want to put emphasis that in the following example, we consider only the query head without taking into account the size of the query body: e.g., by scaling the query path length by a factor of 2, the query head doubles in size; thus, each query transmission requires twice the time. Therefore, since the query has to travel a twice longer path, the QTTR increases by a factor of 4. The four-times increase of QTTR indicates that the relation between the QTTR and the query path length is quadratic.
The quadratic relation can also be observed in the simulation results in Fig.~\ref{fig:qlen}.

Experiment 1 takes into consideration two different network topologies, the GT which resembles an ideal network with constant node density, and the RDT a deployment approach familiar to WSNs where the node density in the network is varying due to the randomized deployment. Surprisingly, in the network deployed following the RDT, the QTTR does not notably diverge from the QTTR in the GT deployment. Therefore, we decided to run experiment 2 to lessen the chance of occasional deployments favoring the RDT and to identify whether the QTTR significantly differs in diverse network topologies. 

The statistical test found no significant differences in QTTR between GT and RDT.
However, in Fig.~\ref{fig:overview}, it is possible to notice that the QTTR in RDT is observed considerably more times outside the interquartile range than in the GT. 
Moreover, in Table~\ref{tab:summary_grid} and Table~\ref{tab:summary_disc}, the max QTTR values are much larger in the RDT which is also subject to aborted queries. In the density chart in Fig.~\ref{fig:topology_distributions} it is possible to notice that density curves of QTTR in GT and RDT are similarly skewed; however, the distribution of QTTR in RDT has two more peaks respectively at 4s and 5s. A possible explanation of these occurrences could be that a network with deployment RDT is more prone to packet loss than a network with GT deployment.

\begin{table*}
\centering
\caption{Summary statistics of the dataset collected in experiment 1, Section~\ref{sec:experiment1} at network deployed according to GT. Description of labels: \textit{\%~aborted queries} the percentage of queries that were aborted due to not reaching the next hop in 30-seconds. \textit{\#one-hop neighborhood} is the number of nodes in the one-hop neighborhood of a node. \textit{\%~reachable nodes} is the percentage of nodes in the same network segment as the sink node.}
\label{tab:summary_grid}
\begin{tabular}{c|c|c|c|c|c|c|c|c|c|ccccc}
\begin{sideways}network size\end{sideways} & \begin{sideways}query path length\end{sideways} & \begin{sideways}query size (bytes)\end{sideways} & \begin{sideways}\textbf{min} QTTR (s)\end{sideways} & \begin{sideways}\textbf{avg} QTTR (s)\end{sideways} & \begin{sideways}\textbf{max} QTTR (s)\end{sideways} & \begin{sideways}\textbf{std} QTTR (s)\end{sideways} & \begin{sideways}\textbf{q25} QTTR (s)\end{sideways} & \begin{sideways}\textbf{median} QTTR (s)\end{sideways} & \begin{sideways}\textbf{q75} QTTR (s)\end{sideways} & \multicolumn{1}{c|}{\begin{sideways}\textbf{\%}~aborted queries\end{sideways}} & \multicolumn{1}{c|}{\begin{sideways}\parbox{2cm}{\textbf{q25 \#}one-hop \\neighborhood}\end{sideways}} & \multicolumn{1}{c|}{\begin{sideways}\parbox{2.3cm}{\textbf{median \#}one-hop \\neighborhood}\end{sideways}} & \multicolumn{1}{c|}{\begin{sideways}\parbox{2cm}{\textbf{q75 \#}one-hop \\neighborhood}\end{sideways}} & \begin{sideways}\textbf{\%}~reachable nodes\end{sideways}  \\
\midrule
50 & 5 & 278 & 0.02 & 0.11 & 1.07 & 0.27 & 0.03 & 0.04 & 0.05 & 0 & 5 & 7 & 8 & 100 \\
50 & 10 & 538 & 0.05 & 0.11 & 1.08 & 0.16 & 0.08 & 0.08 & 0.09 & 0 & & & & \\
50 & 20 & 1058 & 0.17 & 0.23 & 1.21 & 0.16 & 0.2 & 0.21 & 0.22 & 0 & & & & \\
50 & 30 & 1578 & 0.37 & 0.52 & 3.39 & 0.5 & 0.39 & 0.41 & 0.45 & 0 & & & & \\
50 & 40 & 2098 & 0.63 & 0.89 & 1.91 & 0.38 & 0.7 & 0.75 & 0.79 & 0 & & & & \\
50 & 60 & 3138 & 1.5 & 2.66 & 8.82 & 1.52 & 1.7 & 1.95 & 2.86 & 0 & & & & \\
50 & 80 & 4178 & 3.14 & 6.78 & 16.31 & 2.41 & 5.46 & 6.39 & 7.45 & 0 & & & & \\
50 & 100 & 5218 & 11.28 & 19.12 & 31.61 & 5.42 & 15.39 & 17.86 & 21.73 & 0 & & & & \\
\midrule
100 & 5 & 278 & 0.03 & 0.11 & 1.14 & 0.23 & 0.04 & 0.05 & 0.08 & 0 & 5 & 8 & 8 & 100 \\
100 & 10 & 538 & 0.07 & 0.13 & 1.15 & 0.17 & 0.1 & 0.1 & 0.11 & 0 & & & & \\
100 & 20 & 1058 & 0.2 & 0.3 & 1.23 & 0.22 & 0.24 & 0.25 & 0.28 &0 & & & & \\
100 & 30 & 1578 & 0.43 & 0.64 & 3.5 & 0.52 & 0.48 & 0.52 & 0.54 & 0 & & & & \\
100 & 40 & 2098 & 0.71 & 0.91 & 1.88 & 0.18 & 0.82 & 0.89 & 0.94 & 0 & & & & \\
100 & 60 & 3138 & 1.87 & 3.15 & 6.19 & 0.9 & 2.74 & 3.06 & 3.42 & 0 & & & & \\
100 & 80 & 4178 & 3.91 & 8.38 & 14.69 & 2.63 & 6.33 & 8.13 & 10.2 & 0 & & & & \\
100 & 100 & 5218 & 9.67 & 21.17 & 40.76 & 6.32 & 16.49 & 20.77 & 25.16 & 0 & & & & \\
\midrule
200 & 5 & 278 & 0.04 & 0.24 & 2.06 & 0.43 & 0.06 & 0.09 & 0.13 & 0 & 5 & 8 & 8 & 100 \\
200 & 10 & 538 & 0.1 & 0.2 & 1.16 & 0.22 & 0.13 & 0.14 & 0.17 & 0 & & & & \\
200 & 20 & 1058 & 0.27 & 0.42 & 1.36 & 0.31 & 0.3 & 0.32 & 0.35 &0 & & & & \\
200 & 30 & 1578 & 0.5 & 0.69 & 1.69 & 0.32 & 0.56 & 0.58 & 0.64 & 0 & & & & \\
200 & 40 & 2098 & 0.85 & 1.15 & 3.88 & 0.5 & 0.97 & 1.02 & 1.11 & 0 & & & & \\
200 & 60 & 3138 & 2.05 & 3.14 & 7.3 & 1.17 & 2.26 & 3.11 & 3.29 & 0 & & & & \\
200 & 80 & 4178 & 4.89 & 8.69 & 15.21 & 2.41 & 6.89 & 8.24 & 10.41 & 0 & & & & \\
200 & 100 & 5218 & 13.34 & 19.99 & 37.01 & 4.78 & 16.7 & 18.85 & 22.8 & 0 & & & & \\
\midrule
300 & 5 & 278 & 0.05 & 0.31 & 2.09 & 0.47 & 0.08 & 0.13 & 0.18 & 0 & 8 & 8 & 8 & 100 \\
300 & 10 & 538 & 0.13 & 0.28 & 1.22 & 0.3 & 0.16 & 0.18 & 0.21 &0 & & & & \\
300 & 20 & 1058 & 0.28 & 0.48 & 1.41 & 0.35 & 0.32 & 0.36 & 0.41 & 0 & & & & \\
300 & 30 & 1578 & 0.53 & 0.72 & 1.75 & 0.24 & 0.62 & 0.65 & 0.71 & 0 & & & & \\
300 & 40 & 2098 & 0.9 & 1.21 & 2.15 & 0.32 & 1.03 & 1.13 & 1.21 & 0 & & & & \\
300 & 60 & 3138 & 2.25 & 3.53 & 8.81 & 1.21 & 2.56 & 3.36 & 3.86 & 0 & & & & \\
300 & 80 & 4178 & 4.32 & 7.75 & 12.43 & 1.96 & 6.5 & 7.62 & 8.82 & 0 & & & & \\
300 & 100 & 5218 & 11.44 & 17.92 & 30.24 & 3.92 & 15.02 & 18.4 & 20.24 & 0 & & & & \\
\midrule
400 & 5 & 278 & 0.06 & 0.32 & 2.17 & 0.46 & 0.1 & 0.14 & 0.19 & 0 & 8 & 8 & 8 & 100 \\
400 & 10 & 538 & 0.16 & 0.4 & 1.25 & 0.4 & 0.19 & 0.21 & 0.26 &0 & & & & \\
400 & 20 & 1058 & 0.29 & 0.59 & 1.45 & 0.38 & 0.38 & 0.42 & 0.49 & 0 & & & & \\
400 & 30 & 1578 & 0.61 & 0.91 & 3.93 & 0.6 & 0.66 & 0.71 & 0.84 & 0 & & & & \\
400 & 40 & 2098 & 0.96 & 1.38 & 2.4 & 0.4 & 1.13 & 1.2 & 1.4 & 0 & & & & \\
400 & 60 & 3138 & 2.3 & 3.51 & 7.54 & 1.08 & 2.63 & 3.56 & 3.88 & 0 & & & & \\
400 & 80 & 4178 & 4.79 & 8.25 & 12.87 & 1.89 & 6.47 & 8.12 & 9.73 &0 & & & & \\
400 & 100 & 5218 & 12.16 & 18.53 & 27.47 & 3.51 & 15.39 & 18.77 & 20.4 &0 & & & & 
\end{tabular}
\end{table*}

\begin{table*}
\centering
\caption{Summary statistics of the dataset collected in experiment 1, Section~\ref{sec:experiment1} at network deployed according to RDT. Description of labels is given in Table~\ref{tab:summary_grid}.}
\label{tab:summary_disc}
\begin{tabular}{c|c|c|c|c|c|c|c|c|c|ccccc}
\begin{sideways}network size\end{sideways} & \begin{sideways}query path length\end{sideways} & \begin{sideways}query size (bytes)\end{sideways} & \begin{sideways}\textbf{min} QTTR (s)\end{sideways} & \begin{sideways}\textbf{avg} QTTR (s)\end{sideways} & \begin{sideways}\textbf{max} QTTR (s)\end{sideways} & \begin{sideways}\textbf{std} QTTR (s)\end{sideways} & \begin{sideways}\textbf{q25} QTTR (s)\end{sideways} & \begin{sideways}\textbf{median} QTTR (s)\end{sideways} & \begin{sideways}\textbf{q75} QTTR (s)\end{sideways} & \multicolumn{1}{c|}{\begin{sideways}\textbf{\%}~aborted queries\end{sideways}} & \multicolumn{1}{c|}{\begin{sideways}\parbox{2cm}{\textbf{q25 \#}one-hop \\neighborhood}\end{sideways}} & \multicolumn{1}{c|}{\begin{sideways}\parbox{2.3cm}{\textbf{median \#}one-hop \\neighborhood}\end{sideways}} & \multicolumn{1}{c|}{\begin{sideways}\parbox{2cm}{\textbf{q75 \#}one-hop \\neighborhood}\end{sideways}} & \begin{sideways}\textbf{\%}~reachable nodes\end{sideways}\\
 \midrule
50 & 5 & 278 & 0.02 & 0.06 & 1.06 & 0.16 & 0.03 & 0.03 & 0.04 & 0 & 4 & 6 & 10 & 96 \\
50 & 10 & 538 & 0.05 & 0.08 & 0.12 & 0.01 & 0.07 & 0.08 & 0.09 & 0 & & & & \\
50 & 20 & 1058 & 0.15 & 0.21 & 0.3 & 0.03 & 0.19 & 0.21 & 0.22 & 0 & & & & \\
50 & 30 & 1578 & 0.34 & 0.95 & 9.46 & 2.03 & 0.38 & 0.4 & 0.44 & 0 & & & & \\
50 & 40 & 2098 & 0.57 & 1.82 & 9.78 & 2.78 & 0.69 & 0.74 & 0.8 & 0 & & & & \\
50 & 60 & 3138 & 1.32 & 3.79 & 17.6 & 3.98 & 1.57 & 2.12 & 3.47 & 0 & & & & \\
50 & 80 & 4178 & 2.69 & 10.46 & 28.28 & 6.35 & 5.29 & 8.92 & 14.7 & 0 & & & & \\
50 & 100 & 5218 & 8.57 & 31.61 & 76.94 & 14.98 & 19.5 & 28.63 & 43 & 15 & & & & \\
\midrule
100 & 5 & 278 & 0.02 & 0.05 & 0.12 & 0.02 & 0.04 & 0.04 & 0.06 & 0 & 4 & 7 & 10 & 84 \\
100 & 10 & 538 & 0.07 & 0.15 & 1.09 & 0.22 & 0.09 & 0.1 & 0.11 & 0 & & & & \\
100 & 20 & 1058 & 0.19 & 0.36 & 3.26 & 0.52 & 0.21 & 0.23 & 0.25 & 0 & & & & \\
100 & 30 & 1578 & 0.35 & 0.82 & 7.7 & 1.3 & 0.45 & 0.47 & 0.5 & 0 & & & & \\
100 & 40 & 2098 & 0.72 & 1.63 & 9.77 & 2.1 & 0.78 & 0.83 & 0.89 & 0 & & & & \\
100 & 60 & 3138 & 1.62 & 3.46 & 11.91 & 2.89 & 1.83 & 2.1 & 3 & 0 & & & & \\
100 & 80 & 4178 & 3.44 & 10.05 & 24.03 & 5.71 & 5.46 & 7.69 & 14.6 & 0 & & & & \\
100 & 100 & 5218 & 9.58 & 32.38 & 58.73 & 13.05 & 21.4 & 30.45 & 45.88 & 10 & & & & \\
\midrule
200 & 5 & 278 & 0.03 & 0.21 & 1.1 & 0.36 & 0.05 & 0.06 & 0.09 & 0 & 3 & 7 & 12 & 71 \\
200 & 10 & 538 & 0.09 & 0.3 & 4.08 & 0.67 & 0.11 & 0.13 & 0.15 & 0 & & & & \\
200 & 20 & 1058 & 0.21 & 0.7 & 9.28 & 1.46 & 0.27 & 0.28 & 0.32 & 0 & & & & \\
200 & 30 & 1578 & 0.41 & 1.15 & 9.52 & 2.02 & 0.5 & 0.52 & 0.58 & 0 & & & & \\
200 & 40 & 2098 & 0.78 & 1.5 & 9.88 & 1.65 & 0.88 & 0.94 & 1.07 & 0 & & & & \\
200 & 60 & 3138 & 1.68 & 4.07 & 12.75 & 3.17 & 2.06 & 2.93 & 4.13 & 0 & & & & \\
200 & 80 & 4178 & 4.88 & 9.13 & 20.32 & 4.1 & 5.99 & 7.99 & 10.58 & 0 & & & & \\
200 & 100 & 5218 & 15.2 & 30.22 & 80.38 & 13.36 & 21.01 & 28.41 & 35.66 & 17.5 & & & & \\
\midrule
300 & 5 & 278 & 0.04 & 0.17 & 1.11 & 0.28 & 0.06 & 0.08 & 0.1 & 0 & 4 & 6 & 10 & 87 \\
300 & 10 & 538 & 0.1 & 0.26 & 2.2 & 0.39 & 0.14 & 0.15 & 0.17 & 2.5 & & & & \\
300 & 20 & 1058 & 0.27 & 1.06 & 22.33 & 3.49 & 0.33 & 0.37 & 0.4 & 0 & & & & \\
300 & 30 & 1578 & 0.52 & 1.4 & 9.74 & 2.07 & 0.64 & 0.69 & 0.81 & 0 & & & & \\
300 & 40 & 2098 & 0.97 & 1.55 & 10.11 & 1.49 & 1.11 & 1.18 & 1.26 & 0 & & & & \\
300 & 60 & 3138 & 2.19 & 4.02 & 14.39 & 2.34 & 2.6 & 3.4 & 3.93 & 0 & & & & \\
300 & 80 & 4178 & 4.29 & 12.84 & 34.12 & 6.59 & 8.55 & 11 & 14.52 & 0 & & & & \\
300 & 100 & 5218 & 15.59 & 27.55 & 79.09 & 12.76 & 19.48 & 24.34 & 33.05 & 7.5 & & & & \\
\midrule
400 & 5 & 278 & 0.05 & 1.92 & 22.09 & 5.37 & 0.08 & 0.09 & 0.14 & 10 & 4 & 6 & 14 & 86 \\
400 & 10 & 538 & 0.12 & 0.87 & 21.18 & 3.34 & 0.16 & 0.17 & 0.21 & 0 & & & & \\
400 & 20 & 1058 & 0.31 & 0.98 & 9.33 & 1.53 & 0.35 & 0.39 & 1.35 & 2.5 & & & & \\
400 & 30 & 1578 & 0.55 & 1.49 & 9.87 & 2.17 & 0.66 & 0.73 & 0.79 & 12.5 & & & & \\
400 & 40 & 2098 & 0.91 & 1.88 & 7.18 & 1.24 & 1.15 & 1.25 & 2.24 & 0 & & & & \\
400 & 60 & 3138 & 2.15 & 5.88 & 24.57 & 4.95 & 2.88 & 3.72 & 6.49 & 0 & & & & \\
400 & 80 & 4178 & 4.49 & 12.58 & 35.07 & 7.11 & 8.31 & 9.57 & 15.98 & 2.5 & & & & \\
400 & 100 & 5218 & 11.05 & 29.46 & 62.09 & 11.81 & 21 & 28.18 & 34.46 & 10 & & & & 
\end{tabular}
\end{table*}

\section{Conclusion and further work}
\label{sec:conclusion}
In this paper, we propose a technique to retrieve aggregated data from a WSN while preserving data and query privacy. Furthermore, our solution employs the computing power of sensor nodes by conveying them general-purpose computer code for in-situ processing and data aggregation. The proposed solution has been implemented in a simulator, and extensive tests show promising results. 
Analyses suggest the adequacy of the solution in the building monitoring scenario, where the sensor network is deployed on a long-term basis. Moreover, the proposed protocol is generalized and can be used in other networks, where obfuscated computing is required, and computing resources of edge devices can be utilized such as distributed data mining, and federated learning.

However, further studies need to be carried out on this approach, including considerations about unbalanced querying of network regions and active attacks. Active attacks are the ones in which the attacker tries to interfere with the normal functioning of the WSN.


Besides building monitoring, the conceptualized technique can be extended to other scenarios, such as fall prediction in health monitoring using smart floor~\cite{tovsic2021data} or crowd monitoring and detection using WSN~\cite{aggrey2019survey}. We point out a possible application in networks of untrusted devices, in which nodes of the network must jointly compute a function without knowing other parties involved in the computation and without revealing their private inputs. 
Further evaluations on real-life scenarios in a long-time setting will be available in next years as we plan to deploy the proposed technique on a long-time small scale experimental setting in a controlled environment in a cultural heritage building "Mrakova Domačija" in Bled, Slovenia as an extension of experiment presented in~\cite{tovsic2019blockchain}.


%




\ifCLASSOPTIONcaptionsoff
  \newpage
\fi



\bibliographystyle{IEEEtran}
\bibliography{literature}

\begin{thebibliography}{10}
\providecommand{\url}[1]{#1}
\csname url@samestyle\endcsname
\providecommand{\newblock}{\relax}
\providecommand{\bibinfo}[2]{#2}
\providecommand{\BIBentrySTDinterwordspacing}{\spaceskip=0pt\relax}
\providecommand{\BIBentryALTinterwordstretchfactor}{4}
\providecommand{\BIBentryALTinterwordspacing}{\spaceskip=\fontdimen2\font plus
\BIBentryALTinterwordstretchfactor\fontdimen3\font minus
  \fontdimen4\font\relax}
\providecommand{\BIBforeignlanguage}[2]{{%
\expandafter\ifx\csname l@#1\endcsname\relax
\typeout{** WARNING: IEEEtran.bst: No hyphenation pattern has been}%
\typeout{** loaded for the language `#1'. Using the pattern for}%
\typeout{** the default language instead.}%
\else
\language=\csname l@#1\endcsname
\fi
#2}}
\providecommand{\BIBdecl}{\relax}
\BIBdecl

\bibitem{chan2003security}
H.~Chan and A.~Perrig, ``Security and privacy in sensor networks,''
  \emph{computer}, vol.~36, no.~10, pp. 103--105, 2003.

\bibitem{gu2020iotspy}
T.~Gu, Z.~Fang, A.~Abhishek, and P.~Mohapatra, ``Iotspy: Uncovering human
  privacy leakage in iot networks via mining wireless context,'' in \emph{2020
  IEEE 31st Annual International Symposium on Personal, Indoor and Mobile Radio
  Communications}.\hskip 1em plus 0.5em minus 0.4em\relax IEEE, 2020, pp. 1--7.

\bibitem{zhang2011defending}
F.~Zhang, W.~He, and X.~Liu, ``Defending against traffic analysis in wireless
  networks through traffic reshaping,'' in \emph{2011 31st International
  Conference on Distributed Computing Systems}.\hskip 1em plus 0.5em minus
  0.4em\relax IEEE, 2011, pp. 593--602.

\bibitem{saltaformaggio2016eavesdropping}
B.~Saltaformaggio, H.~Choi, K.~Johnson, Y.~Kwon, Q.~Zhang, X.~Zhang, D.~Xu, and
  J.~Qian, ``Eavesdropping on fine-grained user activities within smartphone
  apps over encrypted network traffic,'' in \emph{10th $\{$USENIX$\}$ Workshop
  on Offensive Technologies ($\{$WOOT$\}$ 16)}, 2016.

\bibitem{clements2011sustainable}
D.~Clements-Croome, ``Sustainable intelligent buildings for people: A review,''
  \emph{Intelligent Buildings International}, vol.~3, no.~2, pp. 67--86, 2011.

\bibitem{ghayvat2015enhancement}
H.~Ghayvat, S.~Mukhopadhyay, X.~Gui, and J.~Liu, ``Enhancement of wsn based
  smart home to a smart building for assisted living: Design issues,'' in
  \emph{2015 Fifth International Conference on Communication Systems and
  Network Technologies}.\hskip 1em plus 0.5em minus 0.4em\relax IEEE, 2015, pp.
  219--224.

\bibitem{airQ}
M.~Mrissa, J.~Vcelak, L.~Hajdu, B.~Dávid, M.~Krész, J.~Sandak, A.~Sandak,
  R.~Kanduti, M.~V. Sajn, A.~Jutraz, and K.~M. Rebec, ``Extending {BIM} for air
  quality monitoring,'' in \emph{Construction Materials for a Sustainable
  Future}, ser. {CoMS} 2020/2021, vol.~1.\hskip 1em plus 0.5em minus
  0.4em\relax Slovenian National Building and Civil Engineering Institute,
  2020, pp. 244--250.

\bibitem{henderson2008network}
T.~R. Henderson, M.~Lacage, G.~F. Riley, C.~Dowell, and J.~Kopena, ``Network
  simulations with the ns-3 simulator,'' \emph{SIGCOMM demonstration}, vol.~14,
  no.~14, p. 527, 2008.

\bibitem{ns3}
\BIBentryALTinterwordspacing
{nsnam}, ``ns-3, a discrete-event network simulator for internet systems--
  version 3.32,'' 1-10-2021. [Online]. Available: \url{https://www.nsnam.org/}
\BIBentrySTDinterwordspacing

\bibitem{syverson1997anonymous}
P.~F. Syverson, D.~M. Goldschlag, and M.~G. Reed, ``Anonymous connections and
  onion routing,'' in \emph{Proceedings. 1997 IEEE Symposium on Security and
  Privacy (Cat. No. 97CB36097)}.\hskip 1em plus 0.5em minus 0.4em\relax IEEE,
  1997, pp. 44--54.

\bibitem{pfitzmann2010terminology}
A.~Pfitzmann and M.~Hansen, ``A terminology for talking about privacy by data
  minimization: Anonymity, unlinkability, undetectability, unobservability,
  pseudonymity, and identity management,'' 2010.

\bibitem{ren2010survey}
J.~Ren and J.~Wu, ``Survey on anonymous communications in computer networks,''
  \emph{Computer Communications}, vol.~33, no.~4, pp. 420--431, 2010.

\bibitem{gulcu1996mixing}
C.~Gulcu and G.~Tsudik, ``Mixing e-mail with babel,'' in \emph{Proceedings of
  Internet Society Symposium on Network and Distributed Systems
  Security}.\hskip 1em plus 0.5em minus 0.4em\relax IEEE, 1996, pp. 2--16.

\bibitem{sako1995receipt}
K.~Sako and J.~Kilian, ``Receipt-free mix-type voting scheme,'' in
  \emph{International Conference on the Theory and Applications of
  Cryptographic Techniques}.\hskip 1em plus 0.5em minus 0.4em\relax Springer,
  1995, pp. 393--403.

\bibitem{berthold2001web}
O.~Berthold, H.~Federrath, and S.~K{\"o}psell, ``Web mixes: A system for
  anonymous and unobservable internet access,'' in \emph{Designing privacy
  enhancing technologies}.\hskip 1em plus 0.5em minus 0.4em\relax Springer,
  2001, pp. 115--129.

\bibitem{chaum1981untraceable}
D.~L. Chaum, ``Untraceable electronic mail, return addresses, and digital
  pseudonyms,'' \emph{Communications of the ACM}, vol.~24, no.~2, pp. 84--90,
  1981.

\bibitem{sunshine1977source}
C.~A. Sunshine, ``Source routing in computer networks,'' \emph{ACM SIGCOMM
  Computer Communication Review}, vol.~7, no.~1, pp. 29--33, 1977.

\bibitem{dingledine2004tor}
R.~Dingledine, N.~Mathewson, and P.~Syverson, ``Tor: The second-generation
  onion router,'' Naval Research Lab Washington DC, Tech. Rep., 2004.

\bibitem{li2009privacy}
N.~Li, N.~Zhang, S.~K. Das, and B.~Thuraisingham, ``Privacy preservation in
  wireless sensor networks: A state-of-the-art survey,'' \emph{Ad Hoc
  Networks}, vol.~7, no.~8, pp. 1501--1514, 2009.

\bibitem{jiang2019survey}
J.~Jiang, G.~Han, H.~Wang, and M.~Guizani, ``A survey on location privacy
  protection in wireless sensor networks,'' \emph{Journal of Network and
  Computer Applications}, vol. 125, pp. 93--114, 2019.

\bibitem{carbunar2010query}
B.~Carbunar, Y.~Yu, W.~Shi, M.~Pearce, and V.~Vasudevan, ``Query privacy in
  wireless sensor networks,'' \emph{ACM Transactions on Sensor Networks
  (TOSN)}, vol.~6, no.~2, pp. 1--34, 2010.

\bibitem{xie2017efficient}
K.~Xie, X.~Ning, X.~Wang, S.~He, Z.~Ning, X.~Liu, J.~Wen, and Z.~Qin, ``An
  efficient privacy-preserving compressive data gathering scheme in wsns,''
  \emph{Information Sciences}, vol. 390, pp. 82--94, 2017.

\bibitem{xiang2012compressed}
L.~Xiang, J.~Luo, and C.~Rosenberg, ``Compressed data aggregation:
  Energy-efficient and high-fidelity data collection,'' \emph{IEEE/ACM
  transactions on Networking}, vol.~21, no.~6, pp. 1722--1735, 2012.

\bibitem{donoho2006compressed}
D.~L. Donoho, ``Compressed sensing,'' \emph{IEEE Transactions on information
  theory}, vol.~52, no.~4, pp. 1289--1306, 2006.

\bibitem{li2012compressed}
S.~Li, L.~Da~Xu, and X.~Wang, ``Compressed sensing signal and data acquisition
  in wireless sensor networks and internet of things,'' \emph{IEEE Transactions
  on Industrial Informatics}, vol.~9, no.~4, pp. 2177--2186, 2012.

\bibitem{wu2015efficient}
L.~Wu, K.~Yu, D.~Cao, Y.~Hu, and Z.~Wang, ``Efficient sparse signal
  transmission over a lossy link using compressive sensing,'' \emph{Sensors},
  vol.~15, no.~8, pp. 19\,880--19\,911, 2015.

\bibitem{middya2017compressive}
R.~Middya, N.~Chakravarty, and M.~K. Naskar, ``Compressive sensing in wireless
  sensor networks--a survey,'' \emph{IETE technical review}, vol.~34, no.~6,
  pp. 642--654, 2017.

\bibitem{paillier1999public}
P.~Paillier, ``Public-key cryptosystems based on composite degree residuosity
  classes,'' in \emph{International conference on the theory and applications
  of cryptographic techniques}.\hskip 1em plus 0.5em minus 0.4em\relax
  Springer, 1999, pp. 223--238.

\bibitem{intanagonwiwat2000directed}
C.~Intanagonwiwat, R.~Govindan, and D.~Estrin, ``Directed diffusion: A scalable
  and robust communication paradigm for sensor networks,'' in \emph{Proceedings
  of the 6th annual international conference on Mobile computing and
  networking}, 2000, pp. 56--67.

\bibitem{intanagonwiwat2002impact}
C.~Intanagonwiwat, D.~Estrin, R.~Govindan, and J.~Heidemann, ``Impact of
  network density on data aggregation in wireless sensor networks,'' in
  \emph{Proceedings 22nd international conference on distributed computing
  systems}.\hskip 1em plus 0.5em minus 0.4em\relax IEEE, 2002, pp. 457--458.

\bibitem{tang2006extending}
X.~Tang and J.~Xu, ``Extending network lifetime for precision-constrained data
  aggregation in wireless sensor networks.'' in \emph{INFOCOM}.\hskip 1em plus
  0.5em minus 0.4em\relax Citeseer, 2006, pp. 1--12.

\bibitem{conti2009privacy}
M.~Conti, L.~Zhang, S.~Roy, R.~Di~Pietro, S.~Jajodia, and L.~V. Mancini,
  ``Privacy-preserving robust data aggregation in wireless sensor networks,''
  \emph{Security and Communication Networks}, vol.~2, no.~2, pp. 195--213,
  2009.

\bibitem{zhang2008gp}
W.~Zhang, C.~Wang, and T.~Feng, ``Gp\^{} 2s: Generic privacy-preservation
  solutions for approximate aggregation of sensor data (concise
  contribution),'' in \emph{2008 Sixth Annual IEEE International Conference on
  Pervasive Computing and Communications (PerCom)}.\hskip 1em plus 0.5em minus
  0.4em\relax IEEE, 2008, pp. 179--184.

\bibitem{zhang2013preserving}
L.~Zhang, H.~Zhang, M.~Conti, R.~Di~Pietro, S.~Jajodia, and L.~V. Mancini,
  ``Preserving privacy against external and internal threats in wsn data
  aggregation,'' \emph{Telecommunication Systems}, vol.~52, no.~4, pp.
  2163--2176, 2013.

\bibitem{bista2010privacy}
R.~Bista and J.-W. Chang, ``Privacy-preserving data aggregation protocols for
  wireless sensor networks: a survey,'' \emph{Sensors}, vol.~10, no.~5, pp.
  4577--4601, 2010.

\bibitem{de2009privacy}
E.~De~Cristofaro, X.~Ding, and G.~Tsudik, ``Privacy-preserving querying in
  sensor networks,'' in \emph{2009 Proceedings of 18th International Conference
  on Computer Communications and Networks}.\hskip 1em plus 0.5em minus
  0.4em\relax IEEE, 2009, pp. 1--6.

\bibitem{yick2008wireless}
J.~Yick, B.~Mukherjee, and D.~Ghosal, ``Wireless sensor network survey,''
  \emph{Computer networks}, vol.~52, no.~12, pp. 2292--2330, 2008.

\bibitem{fowler2010domain}
M.~Fowler, \emph{Domain-specific languages}.\hskip 1em plus 0.5em minus
  0.4em\relax Pearson Education, 2010.

\bibitem{castelluccia2005efficient}
C.~Castelluccia, E.~Mykletun, and G.~Tsudik, ``Efficient aggregation of
  encrypted data in wireless sensor networks,'' in \emph{The second annual
  international conference on mobile and ubiquitous systems: networking and
  services}.\hskip 1em plus 0.5em minus 0.4em\relax IEEE, 2005, pp. 109--117.

\bibitem{gao2018mobile}
Y.~Gao, H.~Ao, Z.~Feng, W.~Zhou, S.~Hu, and W.~Tang, ``Mobile network security
  and privacy in wsn,'' \emph{Procedia Computer Science}, vol. 129, pp.
  324--330, 2018.

\bibitem{ieee802}
IEEE, ``Ieee standard for information technology—telecommunications and
  information exchange between systems local and metropolitan area
  networks—specific requirements - part 11: Wireless lan medium access
  control (mac) and physical layer (phy) specifications,'' IEEE, Tech. Rep.,
  2016.

\bibitem{wu2007survey}
B.~Wu, J.~Chen, J.~Wu, and M.~Cardei, ``A survey of attacks and countermeasures
  in mobile ad hoc networks,'' in \emph{Wireless network security}.\hskip 1em
  plus 0.5em minus 0.4em\relax Springer, 2007, pp. 103--135.

\bibitem{shiu2011physical}
Y.-S. Shiu, S.~Y. Chang, H.-C. Wu, S.~C.-H. Huang, and H.-H. Chen, ``Physical
  layer security in wireless networks: A tutorial,'' \emph{IEEE wireless
  Communications}, vol.~18, no.~2, pp. 66--74, 2011.

\bibitem{chandran2001feedback}
K.~Chandran, S.~Raghunathan, S.~Venkatesan, and R.~Prakash, ``A feedback-based
  scheme for improving tcp performance in ad hoc wireless networks,''
  \emph{IEEE Personal communications}, vol.~8, no.~1, pp. 34--39, 2001.

\bibitem{al2005survey}
A.~Al~Hanbali, E.~Altman, and P.~Nain, ``A survey of tcp over ad hoc
  networks.'' \emph{IEEE Commun. Surv. Tutorials}, vol.~7, no. 1-4, pp. 22--36,
  2005.

\bibitem{liu2001atcp}
J.~Liu and S.~Singh, ``Atcp: Tcp for mobile ad hoc networks,'' \emph{IEEE
  Journal on selected areas in communications}, vol.~19, no.~7, pp. 1300--1315,
  2001.

\bibitem{gomez2018tcp}
C.~Gomez, A.~Arcia-Moret, and J.~Crowcroft, ``Tcp in the internet of things:
  from ostracism to prominence,'' \emph{IEEE Internet Computing}, vol.~22,
  no.~1, pp. 29--41, 2018.

\bibitem{clausen2003optimized}
T.~Clausen, P.~Jacquet, C.~Adjih, A.~Laouiti, P.~Minet, P.~Muhlethaler,
  A.~Qayyum, and L.~Viennot, ``Optimized link state routing protocol (olsr),''
  INRIA, RFC, 2003.

\bibitem{abolhasan2004review}
M.~Abolhasan, T.~Wysocki, and E.~Dutkiewicz, ``A review of routing protocols
  for mobile ad hoc networks,'' \emph{Ad hoc networks}, vol.~2, no.~1, pp.
  1--22, 2004.

\bibitem{miller1985use}
V.~S. Miller, ``Use of elliptic curves in cryptography,'' in \emph{Conference
  on the theory and application of cryptographic techniques}.\hskip 1em plus
  0.5em minus 0.4em\relax Springer, 1985, pp. 417--426.

\bibitem{rivest1978method}
R.~L. Rivest, A.~Shamir, and L.~Adleman, ``A method for obtaining digital
  signatures and public-key cryptosystems,'' \emph{Communications of the ACM},
  vol.~21, no.~2, pp. 120--126, 1978.

\bibitem{lara2018elliptic}
C.~A. Lara-Nino, A.~Diaz-Perez, and M.~Morales-Sandoval, ``Elliptic curve
  lightweight cryptography: A survey,'' \emph{IEEE Access}, vol.~6, pp.
  72\,514--72\,550, 2018.

\bibitem{microchip}
K.~Maletsky, ``Rsa vs. ecc comparison for embedded systems,'' Microchip
  Technology Inc., Tech. Rep., 2020.

\bibitem{barkernist}
E.~Barker and Q.~Dang, ``Nist special publication 800-57 part 1, revision 5:
  Recommendation for key management: Part 1--general, may 2020,'' NIST -
  National Institute of Standards and Technology, Tech. Rep., 2020.

\bibitem{NaCi}
``{Libsodium} the sodium crypto library.''
  \url{https://libsodium.gitbook.io/doc/}, accessed: 28.5.2021.

\bibitem{bernstein2006curve25519}
D.~J. Bernstein, ``Curve25519: new diffie-hellman speed records,'' in
  \emph{International Workshop on Public Key Cryptography}.\hskip 1em plus
  0.5em minus 0.4em\relax Springer, 2006, pp. 207--228.

\bibitem{hollander1973wolfe}
D.~A.~W. Myles~Hollander, \emph{Nonparametric Statistical Methods.}\hskip 1em
  plus 0.5em minus 0.4em\relax John Wiley \& Sons, 1973.

\bibitem{galbraith2010study}
S.~Galbraith, J.~A. Daniel, and B.~Vissel, ``A study of clustered data and
  approaches to its analysis,'' \emph{Journal of Neuroscience}, vol.~30,
  no.~32, pp. 10\,601--10\,608, 2010.

\bibitem{brunner2000nonparametric}
E.~Brunner and U.~Munzel, ``The nonparametric behrens-fisher problem:
  Asymptotic theory and a small-sample approximation,'' \emph{Biometrical
  Journal: Journal of Mathematical Methods in Biosciences}, vol.~42, no.~1, pp.
  17--25, 2000.

\bibitem{butun2019security}
I.~Butun, P.~{\"O}sterberg, and H.~Song, ``Security of the internet of things:
  Vulnerabilities, attacks, and countermeasures,'' \emph{IEEE Communications
  Surveys \& Tutorials}, vol.~22, no.~1, pp. 616--644, 2019.

\bibitem{abe2008tag}
M.~Abe, R.~Gennaro, and K.~Kurosawa, ``Tag-kem/dem: A new framework for hybrid
  encryption,'' \emph{Journal of Cryptology}, vol.~21, no.~1, pp. 97--130,
  2008.

\bibitem{tovsic2021data}
A.~To{\v{s}}i{\'c}, N.~Hrovatin, and J.~Vi{\v{c}}i{\v{c}}, ``Data about fall
  events and ordinary daily activities from a sensorized smart floor,''
  \emph{Data in brief}, vol.~37, p. 107253, 2021.

\bibitem{aggrey2019survey}
O.~Aggrey and N.~Evarist, ``Survey of crowd detection algorithms using wireless
  sensor networks: A case of people crowds,'' \emph{International Journal of
  Computer Applications}, vol. 182, no.~38, pp. 1--7, 2019.

\bibitem{tovsic2019blockchain}
A.~To{\v{s}}i{\'c}, J.~Vi{\v{c}}i{\v{c}}, and M.~Mrissa, ``A blockchain-based
  decentralized self-balancing architecture for the web of things,'' in
  \emph{European Conference on Advances in Databases and Information
  Systems}.\hskip 1em plus 0.5em minus 0.4em\relax Springer, 2019, pp.
  325--336.

\end{thebibliography}
\end{document}